\documentclass[12pt]{article}
\usepackage{epsf,amsmath,amssymb,graphicx,scalefnt}
\usepackage{color}
\usepackage{rotating}

\catcode`@=11
\newcount\@tempcntc
\def\@citex[#1]#2{\if@filesw\immediate\write\@auxout{\string\citation{#2}}\fi
  \@tempcnta\z@\@tempcntb\m@ne\def\@citea{}\@cite{\@for\@citeb:=#2\do
    {\@ifundefined
       {b@\@citeb}{\@citeo\@tempcntb\m@ne\@citea\def\@citea{,}{\bf ?}\@warning
       {Citation `\@citeb' on page \thepage \space undefined}}%
    {\setbox\z@\hbox{\global\@tempcntc0\csname b@\@citeb\endcsname\relax}%
     \ifnum\@tempcntc=\z@ \@citeo\@tempcntb\m@ne
       \@citea\def\@citea{,}\hbox{\csname b@\@citeb\endcsname}%
     \else
      \advance\@tempcntb\@ne
      \ifnum\@tempcntb=\@tempcntc
      \else\advance\@tempcntb\m@ne\@citeo
      \@tempcnta\@tempcntc\@tempcntb\@tempcntc\fi\fi}}\@citeo}{#1}}
\def\@citeo{\ifnum\@tempcnta>\@tempcntb\else\@citea\def\@citea{,}%
  \ifnum\@tempcnta=\@tempcntb\the\@tempcnta\else
   {\advance\@tempcnta\@ne\ifnum\@tempcnta=\@tempcntb \else \def\@citea{--}\fi
    \advance\@tempcnta\m@ne\the\@tempcnta\@citea\the\@tempcntb}\fi\fi}
\catcode`@=12

\allowdisplaybreaks
\newcommand{\mhplus}{M_{H^+}}

\title{
  \vskip-3cm{\baselineskip14pt
    \begin{flushleft}
      \normalsize SFB/CPP-12-60 \\
      \normalsize TTP12-29 \\
      \normalsize IFT-5/2012 \\
    \end{flushleft}}
  \vskip1.5cm
  $\bar{B}\to X_s \gamma$ in the Two Higgs Doublet Model\\
   up to Next-to-Next-to-Leading Order in QCD}
\author{Thomas Hermann$^{(a)}$, Miko{\l}aj Misiak$^{(b)}$ and
  Matthias Steinhauser$^{(a)}$\\[2em]
  {\it (a) Institut f\"ur Theoretische Teilchenphysik}\\
  {\it Karlsruhe Institute of Technology (KIT), D-76128 Karlsruhe, Germany}\\[1em]
  {\it (b) Institute of Theoretical Physics, University of Warsaw,}\\
  {\it Ho\.za 69, PL-00-681 Warsaw, Poland}\\
  }

\date{}

\begin{document}
\maketitle


\begin{abstract}

We compute three-loop matching corrections to the Wilson coefficients $C_7$
and $C_8$ in the Two Higgs Doublet Model by applying expansions for small,
intermediate and large charged Higgs boson masses. The results are used to 
evaluate the branching ratio of $\bar{B}\to X_s \gamma$ to next-to-next-to
leading order accuracy, and to determine an updated lower limit on the charged
Higgs boson mass. We find $\mhplus \ge 380\,$GeV at 95\% confidence
level when the recently completed BABAR data analysis is taken into
account. Our results for the charged Higgs contribution to the branching ratio
exhibit considerably weaker sensitivity to the matching scale $\mu_0$, as
compared to previous calculations.

\end{abstract}


\thispagestyle{empty}



\section{Introduction}

In view of missing (to date) New Physics signals at the Large Hadron
Collider (LHC), it is of utmost importance to exploit precision
calculations together with precise experimental results in order to look for
deviations from the Standard Model (SM). In this context, the rare
decay $\bar{B}\to X_s \gamma$ constitutes one of the most important 
processes. It is a loop-generated Flavour-Changing-Neutral-Current (FCNC)
transition, which makes it very sensitive to contributions from beyond-SM
particles. Moreover, its branching ratio can be measured with an
uncertainty of a few percent and, at the same time, the result
can be predicted within perturbation theory with a similar uncertainty.

The current average of the measurements by CLEO~\cite{Chen:2001fj},
BELLE~\cite{Abe:2001hk,Limosani:2009qg} and
BABAR~\cite{Lees:2012ym,Lees:2012wg,Aubert:2007my} reads~\cite{Stone:ICHEP2012}
\begin{eqnarray}
     {\cal B}(\bar{B}\to X_s \gamma)|_{E_\gamma>1.6~\mbox{\tiny GeV}} =
      (3.37 \pm 0.23) \times 10^{-4} \label{exp.aver} 
      \,.
\end{eqnarray}
It includes the recently updated BABAR data
analysis~\cite{Lees:2012ym,Lees:2012wg}.  The measurements have been performed
with various photon energy cutoffs $E_0$ ranging from $1.7$ to
$2.0\,$GeV. Their average in Eq.~(\ref{exp.aver}) involves an extrapolation to
$E_0 = 1.6\,$GeV. It can be confronted with the SM prediction based on the
Next-to-Next-to-Leading Order (NNLO) QCD calculations which reads
${\cal B}(\bar{B}\to X_s \gamma)|_{E_\gamma>1.6\mbox{\scriptsize GeV}}=(3.15
\pm 0.23)\times 10^{-4}$~\cite{Misiak:2006ab,Misiak:2006zs}.

In this paper, we consider extensions of the SM Higgs sector by
a second Higgs doublet, namely the so-called Two Higgs Doublet
Models (2HDMs). They are constructed in such a way that no FCNC occur at
the tree level~\cite{Glashow:1976nt}. Such models
have five physical scalar degrees of freedom, among which there is a charged
Higgs boson $H^\pm$ that plays an important role for $\bar{B}\to X_s
\gamma$. We shall consider two versions of the model, usually denoted
by 2HDM type-I and type-II where, respectively, either
the same or two different Higgs doublet fields couple to the up- and
down-type quarks. In these models, both Higgs doublets acquire vacuum
expectation values $v_{1,2}$ such that $v = \sqrt{v_1^2 + v_2^2} \simeq
246\,$GeV determines the $W^\pm$ and $Z$ boson masses in the same way as in
the SM.  The ratio $v_2/v_1$ is denoted by $\tan\beta$.

Comparison of the experimental results for ${\cal B}(\bar{B}\to X_s \gamma)$
to predictions within the 2HDM type-II leads to the strongest constraint on
the charged Higgs boson mass for $\tan\beta \in [1,25]$ (see, e.g., Sec.~5 of
Ref.~\cite{Flacher:2008zq}; the precise range depends on the treatment of
uncertainties). So far, the constraint has been derived using only the
Next-to-Leading Order (NLO) expressions for the 2HDM contributions at the
electroweak scale, while the SM contributions were treated at the
NNLO~\cite{Misiak:2006zs}. In the present paper, we compute the missing
two- and three-loop NNLO matching coefficients in the 2HDM, and re-do
the analysis to extract a lower bound on $\mhplus$ from the new
experimental average (\ref{exp.aver}).

The outline of the paper is as follows: In the next Section, we discuss the
matching coefficients up to the three-loop order. Besides considering the
2HDM, we also re-compute the SM matching contribution, and improve the
three-loop results for $C_7$ and $C_8$. In Section~\ref{sec::BR2hdm}, we use
the new results to evaluate ${\cal B}(\bar{B}\to X_s \gamma)$ to the
NNLO accuracy, and to extract a lower bound on
$\mhplus$. Section~\ref{sec::conclusions} contains our conclusions.


\section{Matching coefficients}


\subsection{Formalism}

The formalism to compute the Wilson coefficients in the 2HDM can be taken over
from the SM case~\cite{Bobeth:1999mk,Misiak:2004ew}. We shall follow the
regularization and renormalization conventions of those papers. In
particular, we adopt the effective Lagrangian and the definition of the
operators $P_j$ ($j=1,\ldots,8,11$) from Eqs.~(2.1) and~(2.2) of
Ref.~\cite{Misiak:2004ew}. One should note that the dipole operators $P_7$
and $P_8$ are normalized there with inverse powers of the QCD coupling
constant.

Since the additional degrees of freedom of the 2HDM are all heavy, they only
influence the Wilson coefficients $C_i$ of the operators $P_i$. In order to
incorporate the 2HDM contribution in a manner that is analogous to the SM
analysis of Ref.~\cite{Misiak:2004ew}, we split the Wilson coefficients as
\begin{eqnarray}
  C_i^Q &=& C_i^{Q,\rm SM} + C_i^{Q,\rm 2HDM}
  \,,
  \label{eq::CiQ}
\end{eqnarray}
where $Q=c,t$ marks contributions from loop diagrams with virtual charm and
top quarks, respectively.

Our matching calculation is performed at the renormalization scale
$\mu_0$. It is chosen to be of the same order of magnitude as
masses of the
particles that are being decoupled ($m_t$, $M_W$ and $\mhplus$). For the NLO
calculations within the SM we refer to Ref.~\cite{Buras:2002er}.  The NNLO SM
contributions to $C_i^{Q,\rm SM}(\mu_0)$ have been computed in
Refs.~\cite{Bobeth:1999mk,Misiak:2004ew}. Suppression by $m_c^2/M_W^2$ makes
$C_i^{c,\rm 2HDM}$ negligible. In the following, we shall consider $C_i^{t,\rm
2HDM}$ only. It is convenient to decompose them as follows
\begin{eqnarray} C_i^{t,\rm 2HDM} &=&
C_i^{H(0)} + \frac{\alpha_s}{4\pi} C_i^{H(1)} +
\left(\frac{\alpha_s}{4\pi}\right)^2 C_i^{H(2)} +
\left(\frac{\alpha_s}{4\pi}\right)^3 C_i^{H(3)} + \ldots 
\,,
\end{eqnarray} 
where $C_i^{H(n)}$ is obtained from $n$-loop diagrams. For $i=7,8$, the
tree-level coefficients $C_i^{H(0)}$ vanish, while $C_i^{H(1)}(\mu_0)$ and
$C_i^{H(2)}(\mu_0)$ have been found in
Refs.~\cite{Ciuchini:1997xe,Borzumati:1998tg,Ciafaloni:1997un,Bobeth:1999ww}.
%
In the present paper, we compute the three-loop corrections
$C_7^{H(3)}(\mu_0)$ and $C_8^{H(3)}(\mu_0)$, which constitutes the last
missing element of the 2HDM Wilson coefficient evaluation that matters for
${\cal B}(\bar B\to X_s\gamma)$ at the NNLO.  Furthermore, we reproduce the
two-loop corrections $C_i^{H(2)}(\mu_0)$ for $i=3,4,5,6$ that 
have been originally found in Ref.~\cite{Bobeth:2004jz} and 
belong to the necessary NNLO matching in the 2HDM.

In the following two subsections, we discuss our results for the
Wilson coefficients in the SM and 2HDM. Several variables turn out to
be of convenience in this context:
\begin{align}
  y = \frac{M_W}{m_t(\mu_0)}\,,&& 
  w = 1-y^2\,, && 
  r = \frac{m_t^2(\mu_0)}{\mhplus^2}\,,&&
  \bar u = 1- r\,,&&
  u = 1- \frac{1}{r}\,.
\end{align}


\subsection{Standard Model}

Let us begin with recalling three-loop corrections to the matching
coefficients in the SM. They depend on two masses: $M_W$ and $m_t$. 
In Ref.~\cite{Misiak:2004ew}, expansions of the three-loop diagrams 
for $M_W\ll m_t$ and $M_W\approx m_t$ have been performed. The final numerical
results of that paper
\begin{eqnarray}
  C_7^{t(3)}(\mu_0=m_t) &=& 12.05 \pm 0.05\,,\nonumber\\
  C_8^{t(3)}(\mu_0=m_t) &=& -1.2 \pm 0.1\,, 
\label{eq::C78num}
\end{eqnarray}
have been obtained for $y = 0.488 \pm 0.015$. They relied on convergence
behaviour and agreement of the two expansions for 
intermediate values of $M_W$ and $m_t$.  

\begin{figure}[t]
  \begin{center}
    \begin{tabular}{cc}
      \includegraphics[width=.48\textwidth]{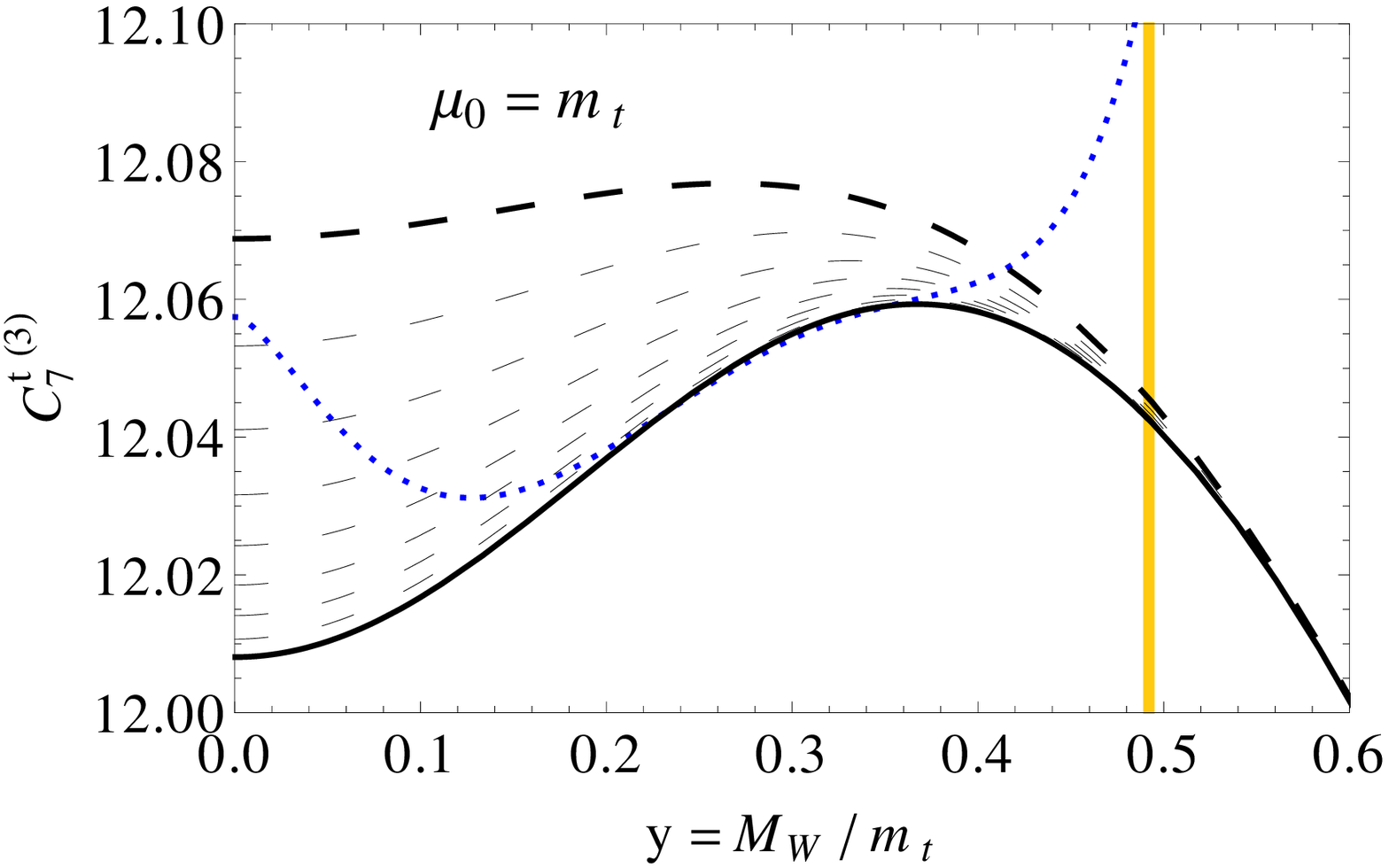} &
      \includegraphics[width=.48\textwidth]{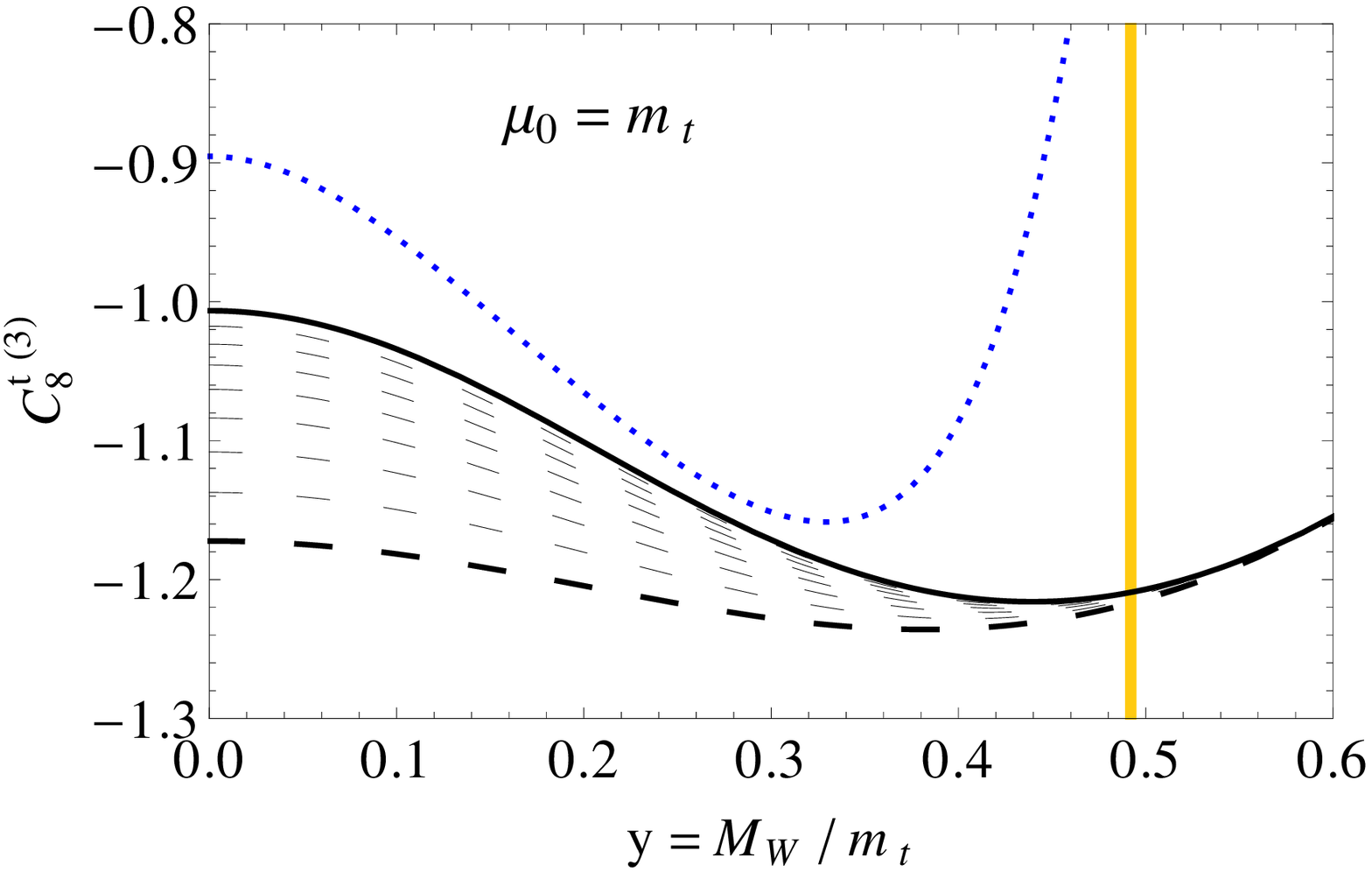}
    \end{tabular}
    \caption{
      \label{fig::c7c8sm}
      Three-loop  SM coefficients $C_7^{t(3)}(\mu_0)$ and $C_8^{t(3)}(\mu_0)$ as
      functions of $y=M_W/m_t$ for $\mu_0=m_t$. Dotted lines show
      their expansions around $y=0$ up to ${\cal O}(y^8)$. Solid lines 
      show the expansions around $y=1$ and include corrections up to 
      $w^{16}= (1-y^2)^{16}$. Dashed lines show lower
      orders in $w$. The (yellow) band represents the physically allowed
      region for $y$.}
  \end{center}
\end{figure}

The dotted curves in Fig.~\ref{fig::c7c8sm} show expansions of
$C_7^{t(3)}(\mu_0=m_t)$ and $C_8^{t(3)}(\mu_0=m_t)$ around $y=0$ up to ${\cal
O}(y^8)$. Results of the expansion around $y =1$ are shown as dashed and solid
lines. The thick dashed lines include terms up to $w^8$ that have been
provided in Ref.~\cite{Misiak:2004ew}.  In the present calculation, we have
added eight more expansion terms for $M_W\approx m_t$. They are shown in
Fig.~\ref{fig::c7c8sm} as thin dashed lines and a solid curve that includes
terms up to $ w^{16}$.

In the case of $C_7^{t(3)}$, we observe an overlap of the two expansions for
$0.2 \leq y \leq 0.35$, which gives us confidence that the exact curve is
approximated with high accuracy by the Taylor expansion around $w=0$ on one
side, and by the asymptotic large-$m_t$ expansion on the other. The situation
for $C_8^{t(3)}$ in Fig.~\ref{fig::c7c8sm}(b) is only slightly worse. We still
observe an improvement w.r.t. Ref.~\cite{Misiak:2004ew} due to the additional
terms in the $w$ expansion.  The vertical bands in 
Fig.~\ref{fig::c7c8sm} correspond to the current experimentally allowed
region $y = 0.492 \pm 0.003$ for $\mu_0 = m_t$. In the range
$0.4<y<0.6$, our improved results are very well approximated (to better
than 0.1\%) by
\begin{eqnarray}
  C_7^{t(3)}(\mu_0=m_t) &=& 11.92 +  0.751\, y -1.03\, y^2
  \,,\nonumber\\
  C_8^{t(3)}(\mu_0=m_t) &=&  -0.764 - 2.06\, y + 2.35\, y^2\,,
\end{eqnarray}
which is consistent with Eq.~(\ref{eq::C78num}) but much more accurate, and
allows to substitute the updated value of $y$. In effect, the uncertainties get
reduced by almost an order of magnitude.

As far as contributions from loops involving the charm quark are concerned,
the corresponding coefficients in the range $0.4<y<0.6$ can be obtained with
high precision from the expressions given already in
Ref.~\cite{Misiak:2004ew}, namely
\begin{eqnarray}
  C_7^{c(3)}(\mu_0=M_W) &=&  1.458\, y^{-0.0338}\,,\nonumber\\
  C_8^{c(3)}(\mu_0=M_W) &=& -1.718\, y^{-0.0598}\,.
\end{eqnarray}
They already provide a high-accuracy approximation in the physical
region, so there is no need to consider higher-order terms in the
expansions.


\subsection{Two Higgs Doublet Model}

In order to specify the notation, we provide the Lagrange density
which defines interactions of the charged Higgs boson with fermions.
Adopting the conventions from Ref.~\cite{Ciuchini:1997xe}, we have
\begin{eqnarray}
  {\cal L}_{H^+} &=& (2\sqrt{2}G_F)^{1/2} \sum_{i,j=1}^3 \overline{u}_i \left(
    A_u m_{u_i} V_{ij} P_L 
    - A_d\, m_{d_j} V_{ij} P_R  \right) d_j H^{+} + h.c.
  \,,
\end{eqnarray}
where $P_{L/R}=(1\mp\gamma_5)/2$, $V_{ij}$ are the Cabibbo-Kobayashi-Maskawa
matrix elements, $u_i$ and $d_j$ are the up- and down-type quarks with masses
$m_{u_i}$ and $m_{d_j}$, and $G_F$ is the Fermi constant.
For the 2HDM of type-I and~II, the couplings $A_d$ and $A_u$ take the 
values
\begin{eqnarray}
  A_u = A_d = \frac{1}{\tan{\beta}} \,
  \label{eq::typeI}
\end{eqnarray}
and
\begin{eqnarray}
 A_u = -\frac{1}{A_d} = \frac{1}{\tan{\beta}} \,,
  \label{eq::typeII}
\end{eqnarray}
respectively.
\begin{figure}[t]
  \begin{center}
    \includegraphics[width=\textwidth]{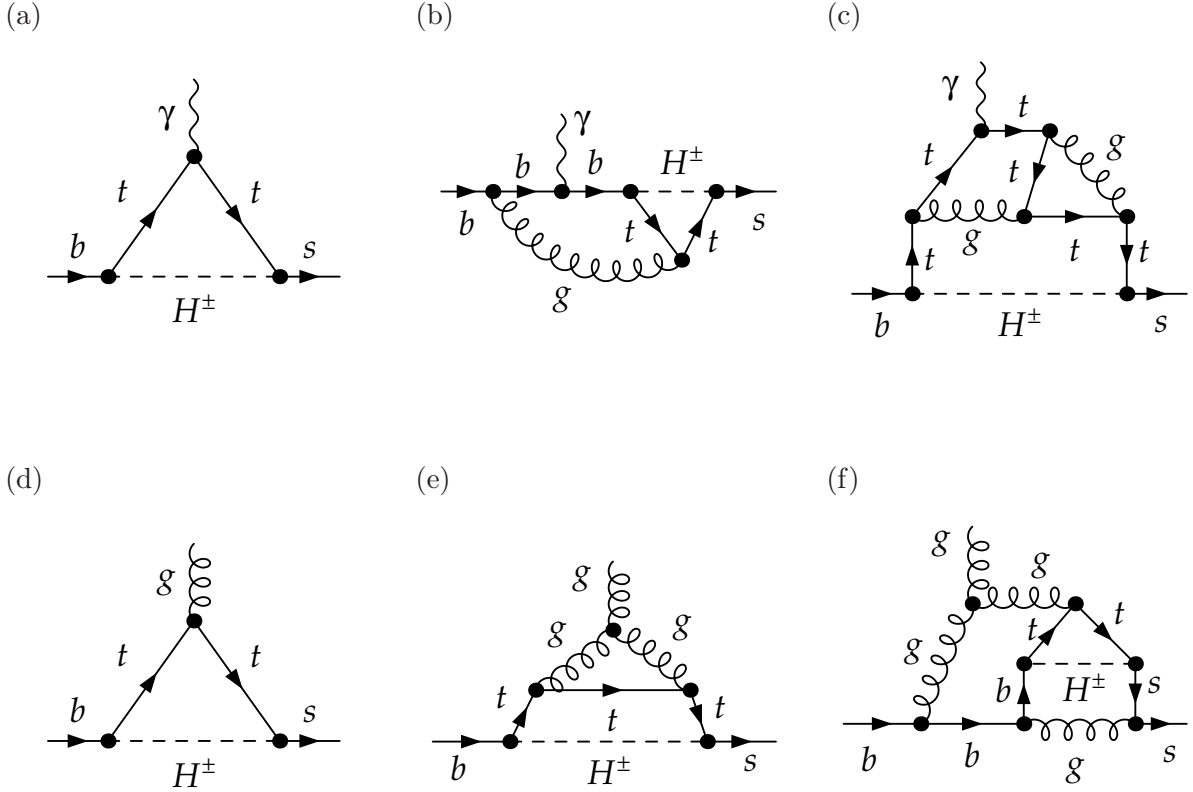}
    \caption{\label{fig::diags} Sample Feynman diagrams contributing to $C_7$
      [(a)-(c)] and $C_8$ [(d)-(f)] at one-, two- and three-loop order.}
  \end{center}
\end{figure}

The 2HDM contributions to the Wilson coefficients in
Eq.~(\ref{eq::CiQ}) are proportional to $A_i A_j^\star$. Since the 
terms involving $A_d^\star$ are suppressed by the strange-quark mass, 
we can safely neglect them. The remaining terms can be split
into two parts as follows:
\begin{eqnarray}
  C_i^{t,\rm 2HDM} = 
  A_d A_u^\star \, C_{i,A_d A_u^{\star}}^{t,\rm 2HDM} 
  +  A_u A_u^\star\, C_{i,A_u A_u^{\star}}^{t,\rm 2HDM} \,.
\end{eqnarray}

For the computation of $C_i^{t,\rm 2HDM}$, we can use Eq.~(5.1) of
Ref.~\cite{Misiak:2004ew} that has been derived in the context of the SM.  Its
application to the 2HDM is straightforward after taking into account that,
apart from the Wilson coefficients, also the electroweak counterterms
(cf. Eqs.~(4.7) to~(4.10) of~\cite{Misiak:2004ew}) and the quantities $B_7$
and $B_8$ (cf. Eqs.~(3.22) and~(3.23) of~\cite{Misiak:2004ew}) receive
additional contributions originating from the charged Higgs boson
exchange. Decomposing each of these quantities as $X = X^{\rm SM} + X^{\rm
2HDM}$ where $X^{\rm SM}$ denotes the result given in
Ref.~\cite{Misiak:2004ew}, we obtain their 2HDM parts in $D=4-2\epsilon$
dimensions by a simple one-loop calculation 
\begin{eqnarray}
  Z_{2,sb}^{t, \rm 2HDM} &=& \frac{m_t^2}{M_W^2} A_u A_u^\star\; \Gamma(\epsilon) 
\left[ -\frac12 + \frac{2r-1}{2(r-1)^2}\left(r^\epsilon-1\right) - \epsilon\,\frac{3r-1}{4(r-1)} 
+ {\cal O}\left(\epsilon^2\right)\right]\,,\nonumber\\[2mm]
  Z_{0,sb}^{t, \rm 2HDM} &=& \frac{\mhplus^2}{M_W^2} A_d A_u^\star\; 
  Z_{0,sb}^{t, \rm SM}|_{M_W \to \mhplus}\,,\nonumber\\[2mm]
  B_k^{\rm 2HDM} &=& A_d A_u^\star\; B_k^{\rm SM}|_{M_W \to \mhplus},\hspace{1cm}
  \mbox{for}\hspace{1cm} k=7,8.
\end{eqnarray}
Both in the SM and 2HDM, the renormalization constants $Z_{0,sb}^Q$ and
$Z_{2,sb}^Q$ enter the electroweak counterterm Lagrangian in the same
way\footnote{
In the corresponding Eq.~(4.6) of Ref.~\cite{Misiak:2004ew}, a factor of
$\frac{M_W^2}{4\pi}e^{\gamma\epsilon}$ is missing in the global normalization,
which we correct here. For clarity, the chirality projectors $P_R$ are now
displayed explicitly.}
\begin{eqnarray} 
{\cal L}^{\rm ew}_{\rm counter} &=& \frac{G_F M_W^2}{4\sqrt{2}\,\pi^2} 
\left[ 
V^*_{cs} V_{cb} \left(\frac{4\pi\mu_0^2}{M_W^2}\right)^\epsilon 
\bar{s} P_R \left( i Z_{2,sb}^c \not\!\!D - Z^c_{0,sb}\, m_b \right) b
\right.\nonumber\\[1mm] && \left. \hspace{13mm} + \; 
V^*_{ts} V_{tb} \left(\frac{4\pi\mu_0^2}{m_t^2}\right)^\epsilon 
\bar{s} P_R \left( i Z_{2,sb}^t \not\!\!D - Z^t_{0,sb}\, m_b \right) b\,
\right], 
\end{eqnarray}
All the remaining quantities appearing in Eq.~(5.1) of Ref.~\cite{Misiak:2004ew}
are precisely the same as in the SM.

In analogy to the SM, we have to consider vacuum integrals with two mass
scales ($\mhplus$ and $m_t$) in our matching calculation. Sample diagrams up
to three-loops are shown in Fig.~\ref{fig::diags}.  At the one- and two-loop
levels, the calculation can be performed exactly, and one obtains 
$C_7$ and $C_8$ as functions of 
$m_t/\mhplus$~\cite{Ciuchini:1997xe,Borzumati:1998tg,Ciafaloni:1997un,Bobeth:1999ww}.
At the three-loop level, we proceed as in Ref.~\cite{Misiak:2004ew},
considering expansions around $m_t\approx \mhplus$, for $m_t\ll \mhplus$, and
for $m_t\gg \mhplus$. In the first case, a simple Taylor expansion is
sufficient, and we have computed the first 16 terms in $u$. Calculations
involving strong hierarchies require non-trivial asymptotic expansions. In
these cases, five terms in $r$ and $1/r$ have been evaluated.

For the purpose of the present analysis, we have re-evaluated the Leading
Order (LO) and NLO contributions to the renormalized Wilson coefficients,
confirming the results of Refs.~\cite{Ciuchini:1997xe,Borzumati:1998tg,Ciafaloni:1997un,Bobeth:1999ww}
and extending them to include higher powers in $\epsilon$. Such an extension
has been necessary for renormalization at the three-loop level. Two-loop
(NNLO) Wilson coefficients of the four-quark-operators were calculated
in Ref.~\cite{Bobeth:2004jz} for the MSSM. We have performed an independent
calculation of the charged Higgs boson contribution. Full agreement has been
found.  We refrain from displaying here explicit analytical results for the
one- and two-loop Wilson coefficients, and refer to the {\tt Mathematica} file
available from Ref.~\cite{progdata}. Let us only note that we set $m_t =
m_t(\mu_0)$ in all those lower-order terms, which specifies our conventions
for the NNLO expressions below.

At the three-loop level, we have been able to recover the dependence of 
$C_7$ and $C_8$ on $\mu_0$ by applying renormalization group techniques to the
analytical one- and two-loop results. We find 
\begin{eqnarray}
  C_{7,A_u A_u^{\star}}^{H(3)}(\mu_0) &=& C_{7,A_u A_u^{\star}}^{H(3)}(\mu_0=m_t)\nonumber\\
&&  + \ln\left(\frac{\mu_0^2}{m_t^2}\right) \left[ -\frac{r \left(67930 r^4-470095 r^3+1358478 r^2-700243 r+54970\right)}{2187 (r-1)^5} \right. \nonumber\\
&& + \frac{r \left(10422 r^4-84390 r^3+322801 r^2-146588 r+1435\right)}{729 (r-1)^6} \ln r \nonumber\\
&& \left. + \frac{2 r^2 \left(260 r^3-1515 r^2+3757 r-1446\right)}{27
    (r-1)^5} 
  \text{Li}_2\left(1-\frac{1}{r}\right) \right] \nonumber\\
&&  + \ln^2\left(\frac{\mu_0^2}{m_t^2}\right) \left[ \frac{r \left(-518 r^4+3665 r^3-17397 r^2+3767 r+1843\right)}{162 (r-1)^5} \right. \nonumber\\
&& \left.+ \frac{r^2 \left(-63 r^3+532 r^2+2089 r-1118\right)}{27 (r-1)^6} \ln r \right] 
  \,,\\
C_{7,A_d A_u^{\star}}^{H(3)}(\mu_0) &=& C_{7,A_d A_u^{\star}}^{H(3)}(\mu_0=m_t) \nonumber\\
&& + \ln\left(\frac{\mu_0^2}{m_t^2}\right) \left[ \frac{r \left(3790 r^3-22511 r^2+53614 r-21069\right)}{81 (r-1)^4} \right.\nonumber\\
&& + \frac{2 r \left(-1266 r^3+7642 r^2-21467 r+8179\right)}{81 (r-1)^5} \ln r \nonumber\\
&& \left. -\frac{8 r \left(139 r^3-612 r^2+1103 r-342\right)}{27
    (r-1)^4} 
  \text{Li}_2\left(1-\frac{1}{r}\right) \right] \nonumber\\
&&  + \ln^2\left(\frac{\mu_0^2}{m_t^2}\right) \left[ \frac{r \left(284 r^3-1435 r^2+4304 r-1425\right)}{27 (r-1)^4} \right.\nonumber\\
&& \left. + \frac{2 r \left(63 r^3-397 r^2-970 r+440\right)}{27 (r-1)^5} \ln r \right] 
  \,,\\
  C_{8,A_u A_u^{\star}}^{H(3)}(\mu_0) &=& C_{8,A_u A_u^{\star}}^{H(3)}(\mu_0=m_t)\nonumber\\ 
&&  
+ \ln\left(\frac{\mu_0^2}{m_t^2}\right) \left[ 
  \frac{r \left(51948 r^4-233781 r^3+48634 r^2-698693 r+2452\right)}{1944
  (r-1)^6} \ln r
\right.\nonumber\\
&& 
-\frac{r \left(522347
      r^4-2423255 r^3+2706021 r^2-5930609 r+148856\right)}{11664 (r-1)^5} 
\nonumber\\
&& \left. + \frac{r^2 \left(481 r^3-1950 r^2+1523 r-2550\right)}{18
    (r-1)^5} 
  \text{Li}_2\left(1-\frac{1}{r}\right) \right]\nonumber\\
&&  + \ln^2\left(\frac{\mu_0^2}{m_t^2}\right) \left[ \frac{r \left(-259 r^4+1117 r^3+2925 r^2+28411 r+2366\right)}{216 (r-1)^5} \right. \nonumber\\
&&\left. -\frac{r^2 \left(139 r^2+2938 r+2683\right)}{36 (r-1)^6} \ln r \right] 
  \,,\\
C_{8,A_d A_u^{\star}}^{H(3)}(\mu_0) &=& C_{8,A_d A_u^{\star}}^{H(3)}(\mu_0=m_t) 
  + \ln\left(\frac{\mu_0^2}{m_t^2}\right) \left[ \frac{r \left(1463 r^3-5794 r^2+5543 r-15036\right)}{27 (r-1)^4} \right. \nonumber\\
&& + \frac{r \left(-1887 r^3+7115 r^2+2519 r+19901\right)}{54 (r-1)^5} \ln r \nonumber\\
&&\left. + \frac{r \left(-629 r^3+2178 r^2-1729 r+2196\right)}{18
    (r-1)^4} 
  \text{Li}_2\left(1-\frac{1}{r}\right) \right] \nonumber \\
&&  + \ln^2\left(\frac{\mu_0^2}{m_t^2}\right) \left[ \frac{r \left(259 r^3-947 r^2-251 r-5973\right)}{36 (r-1)^4}\right.\nonumber\\
&& \left. + \frac{r \left(139 r^2+2134 r+1183\right)}{18 (r-1)^5} \ln r \right] 
  \,.
\end{eqnarray}

Our results for $C_7^{H(3)}(\mu_0=m_t)$ and $C_8^{H(3)}(\mu_0=m_t)$ 
in terms of expansions are quite lengthy. Exact values of the expansion
coefficients can be found in Ref.~\cite{progdata}. Here, we present
their approximate numerical values only. Considering consecutively the 
regions $r\to0$, $r\to1$ and $r\to\infty$, we obtain (for $\mu_0=m_t$):
\begin{eqnarray}
C_{7,A_u A_u^{\star}}^{H(3),r\to 0} &=& 
    0.9225\, r \ln^2 r +4.317\, r \ln r -8.278\, r \nonumber\\
&& -20.73\, r^2 \ln^3 r-112.4\, r^2 \ln^2 r -396.1\, r^2 \ln r -480.9\, r^2 \nonumber\\
&& -34.50\, r^3 \ln^3 r - 348.2\, r^3 \ln^2 r - 1292\, r^3 \ln r -1158\, r^3 \nonumber\\
&& -23.26\, r^4 \ln^3 r -541.4\, r^4 \ln^2 r -2540\, r^4 \ln r -1492 \, r^4 \nonumber\\
&& +42.30\, r^5 \ln^3 r-412.4\, r^5 \ln^2 r - 3362\, r^5 \ln r-823.0\, r^5 
   + {\cal O}\left(r^6\right)\,,\\[2mm]
C_{7,A_u A_u^{\star}}^{H(3),r\to 1^-} &=& 
     1.283 - 0.7158 \, \bar u - 0.3039 \, \bar u^2 - 0.1549 \, \bar u^3 - 0.08625 \, \bar u^4 - 0.05020 \, \bar u^5 \nonumber\\
&& - 0.02970 \, \bar u^6 - 0.01740 \, \bar u^7 - 0.009752 \, \bar u^8 - 0.004877 \, \bar u^9 \nonumber\\
&& - 0.001721 \, \bar u^{10} + 0.0003378 \, \bar u^{11} + 0.001679 \, \bar u^{12} + 0.002542 \, \bar u^{13} \nonumber\\
&& + 0.003083 \, \bar u^{14} + 0.003404 \, \bar u^{15} + 0.003574 \, \bar u^{16} 
   + {\cal O}\left(\bar u^{17}\right)\,,\\[2mm]
C_{7,A_u A_u^{\star}}^{H(3),r\to 1^+} &=& 
      1.283 + 0.7158\, u + 0.4119\, u^2 + 0.2629\, u^3 + 0.1825\, u^4 + 0.1347\, u^5 \nonumber\\
&&  + 0.1040\, u^6 + 0.08306\, u^7 + 0.06804\, u^8 + 0.05688\, u^9 + 0.04833\, u^{10} \nonumber\\
&& + 0.04163\, u^{11} + 0.03625\, u^{12} + 0.03188\, u^{13} + 0.02827\, u^{14} + 0.02525\, u^{15} \nonumber\\ 
&& + 0.02269\, u^{16} + {\cal O}\left(u^{17}\right)\,,\\[2mm]
C_{7,A_u A_u^{\star}}^{H(3),r\to\infty} &=& 
     3.970 - 8.753 \frac{\ln r}{r} + 15.35 \frac{1}{r} - 38.12 \frac{\ln r}{r^2} + 47.09 \frac{1}{r^2} - 103.8 \frac{\ln r}{r^3}\nonumber\\
&& + 79.15 \frac{1}{r^3} - 168.3 \frac{\ln r}{r^4} + 24.41\frac{1}{r^4} - 72.13 \frac{\ln r}{r^5} -274.2\frac{1}{r^5} 
   + {\cal O}\left(\frac{1}{r^6}\right)\!,\\[2mm]
C_{7,A_d A_u^{\star}}^{H(3),r\to 0} &=& 
    -20.94 \, r \ln^3 r - 123.5 \, r \ln^2 r - 453.5 \, r \ln r - 572.2 \, r \nonumber\\
&& - 8.889 \, r^2 \ln^3 r - 195.7\, r^2 \ln^2 r - 870.3 \, r^2 \ln r - 524.1\, r^2\nonumber\\
&& + 19.73\, r^3 \ln^3 r - 46.61 \, r^3 \ln^2 r - 826.2 \, r^3 \ln r + 166.7 \, r^3\nonumber\\
&& + 36.08 \, r^4 \ln^3 r + 323.2\, r^4 \ln^2 r + 169.9 \, r^4 \ln r + 1480 \, r^4\nonumber\\
&& - 66.63\, r^5 \ln^3 r + 469.4\, r^5 \ln^2 r + 1986\, r^5 \ln r + 2828\, r^5 
   + {\cal O}\left(r^6\right)\,,\\[2mm]
C_{7,A_d A_u^{\star}}^{H(3),r\to 1^-} &=& 
     12.82 + 1.663\, \bar u + 0.7780 \, \bar u^2 + 0.3755 \, \bar u^3 + 0.1581 \, \bar u^4\nonumber\\
&& + 0.03021\, \bar u^5 - 0.04868 \, \bar u^6 - 0.09864 \, \bar u^7 - 0.1306 \, \bar u^8\nonumber\\
&& - 0.1510 \, \bar u^9 - 0.1637 \, \bar u^{10} - 0.1712 \, \bar u^{11} - 0.1751 \, \bar u^{12}\nonumber\\
&& - 0.1766\, \bar u^{13} - 0.1763 \, \bar u^{14} - 0.1748 \, \bar u^{15} - 0.1724 \, \bar u^{16} 
   + {\cal O}\left(\bar u^{17}\right)\,,\\[2mm]
C_{7,A_d A_u^{\star}}^{H(3),r\to 1^+} &=& 
    12.82 - 1.663\, u - 0.8852\, u^2 - 0.4827\, u^3 - 0.2976\, u^4 - 0.2021\, u^5\nonumber\\
&&  - 0.1470\, u^6 - 0.1125\, u^7 - 0.08931\, u^8 - 0.07291\, u^9 - 0.06083\, u^{10}\nonumber\\
&&  - 0.05164\, u^{11} - 0.04446\, u^{12} - 0.03873\, u^{13} - 0.03407\, u^{14} - 0.03023\, u^{15}\nonumber\\ 
&&  - 0.02702\, u^{16} + {\cal O}\left(u^{17}\right)\,,\\[2mm]
C_{7,A_d A_u^{\star}}^{H(3),r\to\infty} &=& 
     8.088 + 9.757\frac{\ln r}{r} - 12.91\frac{1}{r} + 38.43\frac{\ln r}{r^2} -49.32\frac{1}{r^2} + 106.2\frac{\ln r}{r^3}\nonumber\\
&& - 78.90\frac{1}{r^3} + 168.4\frac{\ln r}{r^4} - 24.97 \frac{1}{r^4} + 101.1 \frac{\ln r}{r^5}+194.3\frac{1}{r^5}
   + {\cal O}\left(\frac{1}{r^6}\right),\\[2mm]
C_{8,A_u A_u^{\star}}^{H(3),r\to 0} &=& 
     0.6908 \, r \ln^2 r + 3.238 \, r \ln r + 0.7437  \, r \nonumber\\
&& - 22.98 \, r^2 \ln^3 r - 169.1 \, r^2 \ln^2 r - 602.7 \, r^2 \ln r - 805.5 \, r^2 \nonumber\\
&& - 66.32 \, r^3 \ln^3 r - 779.6 \, r^3 \ln^2 r - 3077 \, r^3 \ln r - 3357 \, r^3 \nonumber\\
&& - 143.4 \, r^4 \ln^3 r - 2244 \, r^4 \ln^2 r - 10102 \, r^4 \ln r - 9016 \, r^4 \nonumber\\
&& - 226.7 \, r^5 \ln^3 r - 5251 \, r^5 \ln^2 r - 26090 \, r^5 \ln r -19606\, r^5  
   + {\cal O}\left(r^6\right)\,,\\[2mm]
C_{8,A_u A_u^{\star}}^{H(3),r\to 1^-} &=& 
     1.188 - 0.4078 \, \bar u - 0.2076 \, \bar u^2 - 0.1265 \, \bar u^3 - 0.08570 \, \bar u^4 - 0.06204 \, \bar u^5 \nonumber\\
&& - 0.04689 \, \bar u^6 - 0.03652 \, \bar u^7 - 0.02907 \, \bar u^8 - 0.02354 \, \bar u^9 \nonumber\\
&& - 0.01933 \, \bar u^{10} - 0.01605 \, \bar u^{11} - 0.01345 \, \bar u^{12}\nonumber - 0.01137 \, \bar u^{13} \\
&& - 0.009678 \, \bar u^{14} - 0.008293 \, \bar u^{15} - 0.007148 \, \bar u^{16} 
   + {\cal O}\left(\bar u^{17}\right)\,,\\[2mm]
C_{8,A_u A_u^{\star}}^{H(3),r\to 1^+} &=& 
    1.188 + 0.4078\, u + 0.2002\, u^2 + 0.1190\, u^3 + 0.07861\, u^4 + 0.05531\, u^5\nonumber\\
&&  + 0.04061\, u^6 + 0.03075\, u^7 + 0.02386\, u^8 + 0.01888\, u^9 + 0.01520\, u^{10}\nonumber\\
&&  + 0.01241\, u^{11} + 0.01026\, u^{12} + 0.008575\, u^{13} + 0.007238\, u^{14}\nonumber\\
&&  + 0.006164\, u^{15} + 0.005290\, u^{16} + {\cal O}\left(u^{17}\right)\,,\\[1mm]
C_{8,A_u A_u^{\star}}^{H(3),r\to\infty} &=& 
     2.278 - 5.214 \frac{1}{r} + 20.02 \frac{\ln r}{r^2} - 39.76 \frac{1}{r^2} + 78.58 \frac{\ln r}{r^3} - 66.39 \frac{1}{r^3}\nonumber\\
&& + 91.89 \frac{\ln r}{r^4} + 96.35 \frac{1}{r^4} - 300.7 \frac{\ln r}{r^5} + 826.2 \frac{1}{r^5}
   + {\cal O}\left(\frac{1}{r^6}\right)\,,\\[2mm]
C_{8,A_d A_u^{\star}}^{H(3),r\to 0} &=& 
   - 19.80 \, r \ln^3 r - 174.7 \, r \ln^2 r - 658.4 \, r \ln r - 929.8 \, r \nonumber\\
&& - 31.83 \, r^2 \ln^3 r - 612.6 \, r^2 \ln^2 r - 2770 \, r^2 \ln r - 2943 \, r^2 \nonumber\\
&& - 40.68 \, r^3 \ln^3 r - 1439 \, r^3 \ln^2 r - 7906 \, r^3 \ln r - 6481 \, r^3 \nonumber\\
&& + 54.66 \, r^4 \ln^3 r - 2777 \, r^4 \ln^2 r - 17770 \, r^4 \ln r - 11684 \, r^4\nonumber\\
&& + 1003\, r^5 \ln^3 r - 2627 \, r^5 \ln^2 r - 29962\, r^5 \ln r -15962\, r^5 
   + {\cal O}\left(r^6\right)\,,\\[2mm]
C_{8,A_d A_u^{\star}}^{H(3),r\to 1^-} &=& 
    -0.6110 + 1.095 \, \bar u + 0.6492 \, \bar u^2 + 0.4596 \, \bar u^3 + 0.3569 \, \bar u^4\nonumber\\
&& + 0.2910 \, \bar u^5 + 0.2438 \, \bar u^6 + 0.2075 \, \bar u^7 + 0.1785 \, \bar u^8\nonumber\\
&& + 0.1546 \, \bar u^9 + 0.1347 \, \bar u^{10} + 0.1177 \, \bar u^{11} + 0.1032 \, \bar u^{12}\nonumber\\
&& + 0.09073 \, \bar u^{13} + 0.07987 \, \bar u^{14} + 0.07040 \, \bar u^{15} + 0.06210 \, \bar u^{16} 
   + {\cal O}\left(\bar u^{17}\right)\,,\\[2mm]
C_{8,A_d A_u^{\star}}^{H(3),r\to 1^+} &=& 
    -0.6110 - 1.095\, u - 0.4463\, u^2 - 0.2568\, u^3 - 0.1698\, u^4 - 0.1197\, u^5 \nonumber\\
&&  - 0.08761\, u^6 - 0.06595\, u^7 - 0.05079\, u^8 - 0.03987\, u^9 - 0.03182\, u^{10} \nonumber\\
&&  - 0.02577\, u^{11} - 0.02114\, u^{12} - 0.01754\, u^{13} - 0.01471\, u^{14} - 0.01244\, u^{15} \nonumber\\ 
&&  - 0.01062\, u^{16} + {\cal O}\left(u^{17}\right)\,,\\[2mm]
C_{8,A_d A_u^{\star}}^{H(3),r\to\infty} &=& 
   - 3.174 + 10.89 \frac{1}{r} - 35.42 \frac{\ln r}{r^2} + 63.74 \frac{1}{r^2} -110.7 \frac{\ln r}{r^3} + 62.26 \frac{1}{r^3}\nonumber\\
&& - 71.62 \frac{\ln r}{r^4} -205.7 \frac{1}{r^4} + 476.9 \frac{\ln r}{r^5} -1003 \frac{1}{r^5}
   + {\cal O}\left(\frac{1}{r^6}\right)\,.
\end{eqnarray}
\begin{figure}[t]
  \begin{center}
    \begin{tabular}{cc}
      \includegraphics[width=0.48\textwidth]{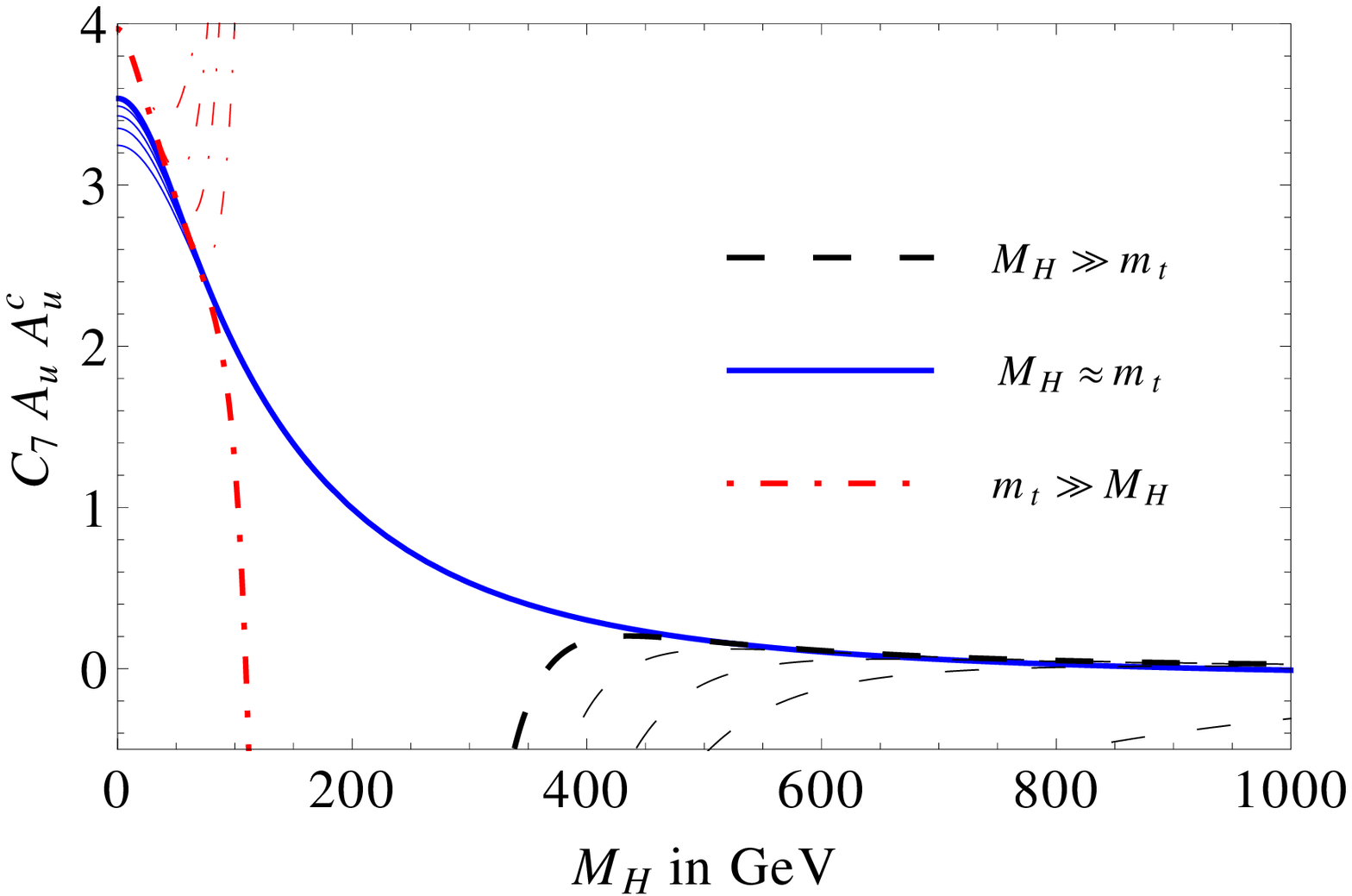} &
      \includegraphics[width=0.48\textwidth]{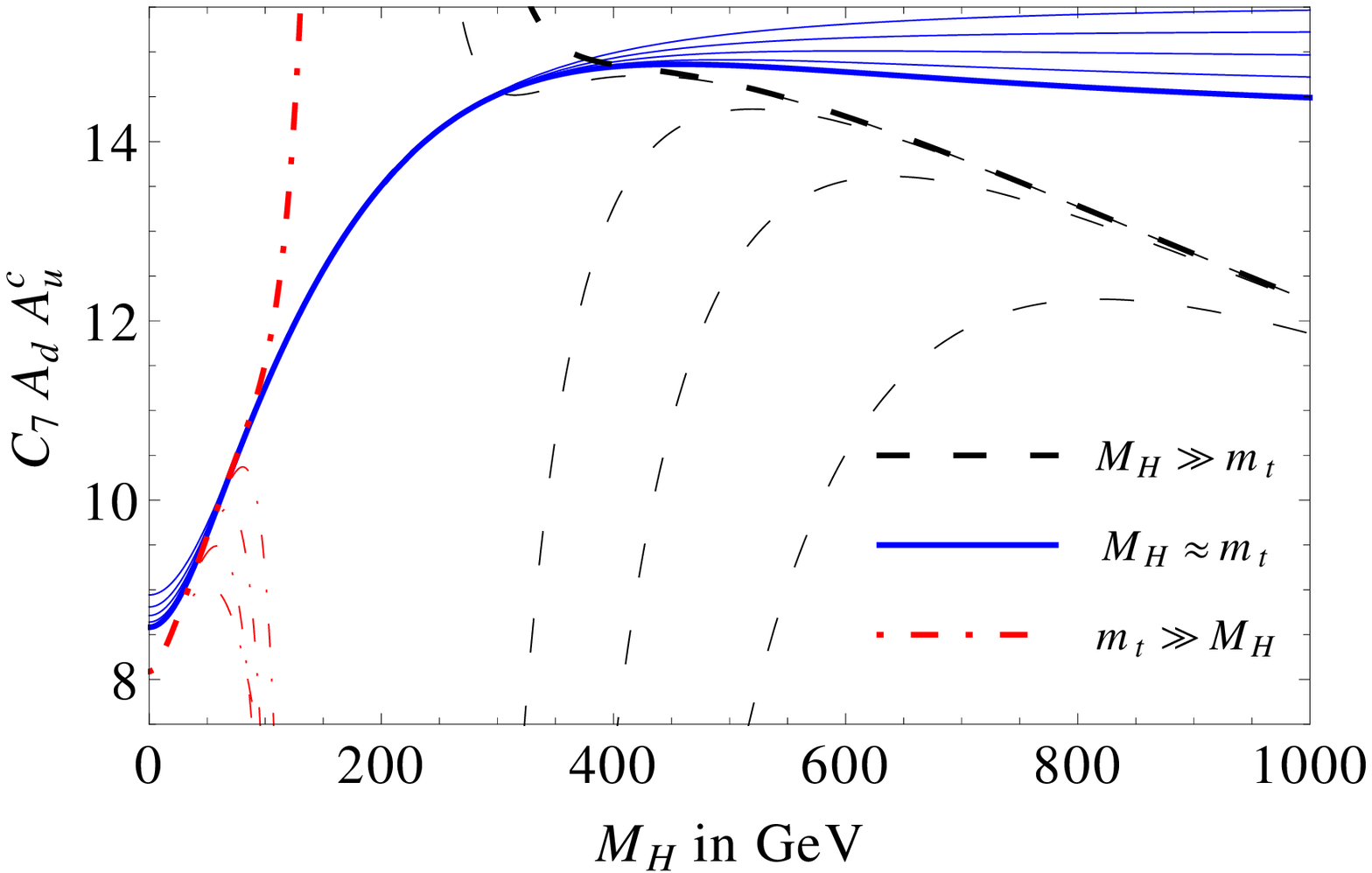}
      \\
      (a) & (b)
    \end{tabular}
    \caption{
      \label{fig::c73l2hdm}
      Three-loop coefficients 
      $C_{7,A_u A_u^{\star}}^{H(3)}(\mu_0=m_t)$~~(a)
      and 
      $C_{7,A_d A_u^{\star}}^{H(3)}(\mu_0=m_t)$~~(b) 
      as functions of $\mhplus$. The dashed, solid and dash-dotted lines
      correspond to the expansions for $\mhplus\to\infty$, $\mhplus\approx m_t$
      and $\mhplus\to0$, respectively.}
  \end{center}
\end{figure}

At this point, a comment concerning the expansions in $\bar
u=1-m_t^2(\mu_0)/\mhplus^2$ and $u=1-\mhplus^2/m_t^2(\mu_0)$ is in
order.  For $\mhplus\approx m_t$, one can expand either in $\bar u$ or in $u$,
and the two expansions are easily convertible into each other.\footnote{
The exact expansion coefficients from Ref.~\cite{progdata} should be used for the conversion.}
As it has been observed in Ref.~\cite{Eiras:2006xm}, one expects better
convergence of the expansion for $\mhplus\ge m_t$ ($\mhplus\le m_t$) if the
result is expressed in terms of $\bar u$ ($u$).  In the following, we
shall always choose the better-suited representation without explicitly
mentioning it.
\begin{figure}[t]
\begin{center}
  \begin{tabular}{cc}
    \includegraphics[width=0.48\textwidth]{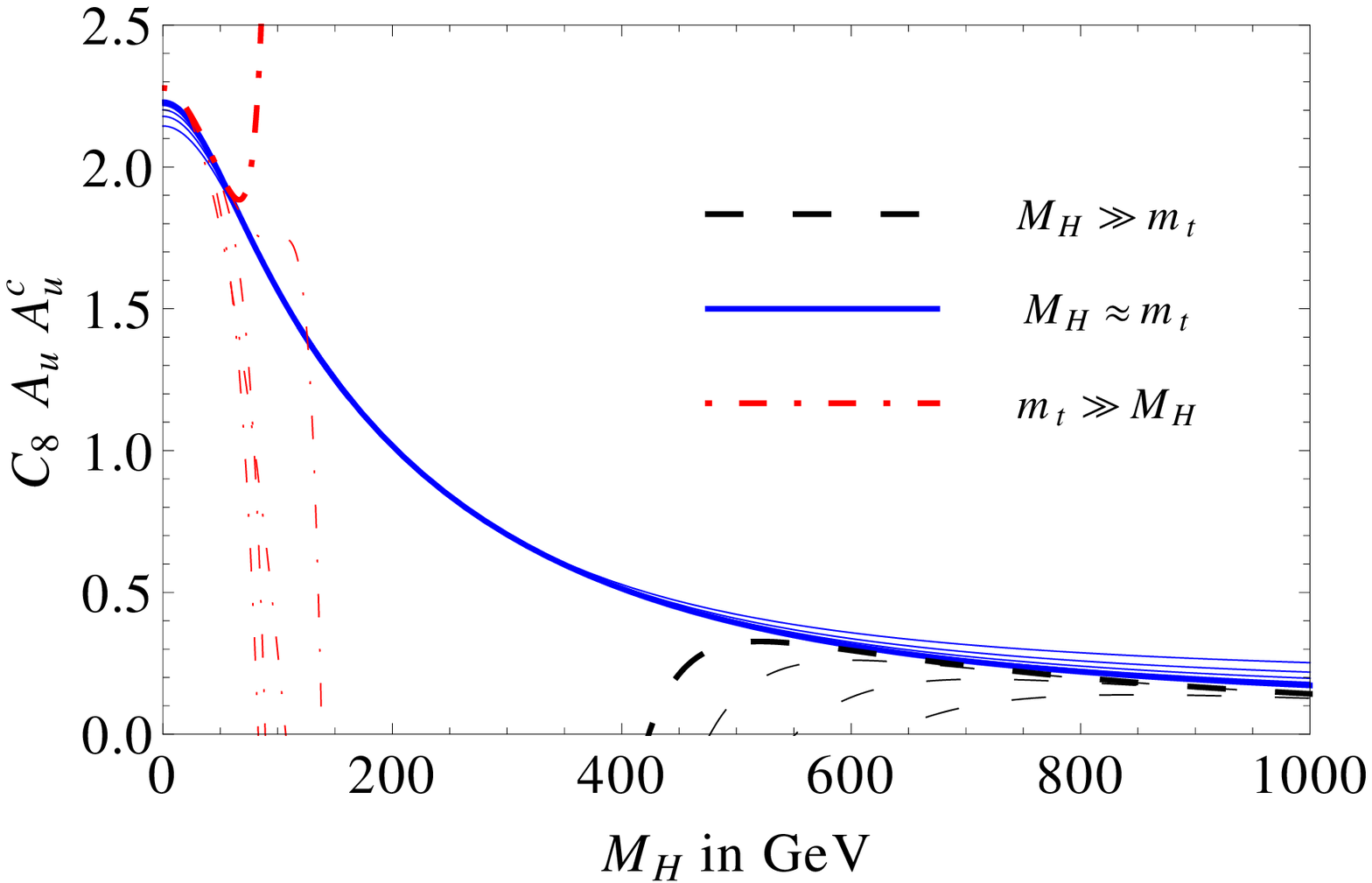} &
    \includegraphics[width=0.48\textwidth]{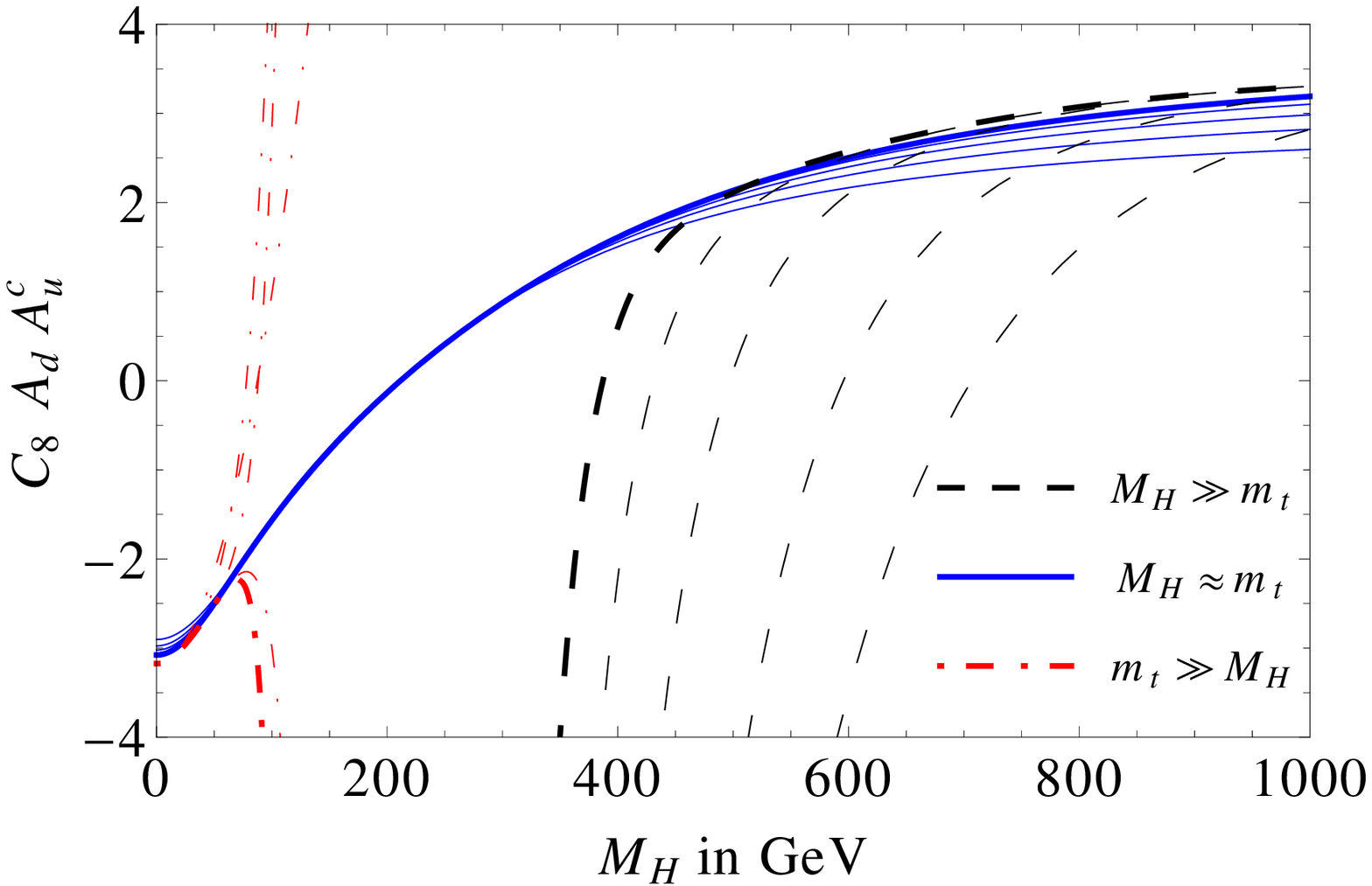}
    \\
    (a) & (b)
  \end{tabular}
  \caption{
    \label{fig::c83l2hdm}
      Three-loop coefficients 
      $C_{8,A_u A_u^{\star}}^{H(3)}(\mu_0=m_t)$~~(a)
      and 
      $C_{8,A_d A_u^{\star}}^{H(3)}(\mu_0=m_t)$~~(b) 
      as functions of $\mhplus$. The dashed, solid and dash-dotted lines
      correspond to the expansions for $\mhplus\to\infty$, $\mhplus\approx m_t$
      and $\mhplus\to0$, respectively.}
\end{center}
\end{figure}

In Figs.~\ref{fig::c73l2hdm} and~\ref{fig::c83l2hdm}, we demonstrate that the
above expansions are sufficient to obtain the Wilson coefficients for any
$\mhplus$. In Fig.~\ref{fig::c73l2hdm}(a), the coefficient $C_{7,A_u
A_u^{\star}}^{H(3)}(\mu_0=m_t)$ is plotted as a function of $\mhplus$.  The
thick-dashed, solid and dash-dotted lines show the results for $r\to0$,
$r\to1$ and $r\to\infty$, respectively, including the highest available
expansion coefficients.  Convergence of the expansions is illustrated by the
thin lines that describe lower orders in the respective expansions. One
observes that the thick dash-dotted and solid curves overlap for
$\mhplus\approx 30 -70\,$GeV, which suggests that good approximations to the
(unknown) exact results are provided by the $r\to\infty$ and $r\to 1$
expansions for $\mhplus<50\,$GeV and $\mhplus>50\,$GeV, respectively.
Similarly, for $\mhplus\approx 500 - 650\,$GeV, one observes agreement between
the solid and dashed curves, which justifies the use of the $r\to1$ result for
$\mhplus<520\,$GeV and the $r\to0$ result above this value. In this way, we
can define $C_{7,A_u A_u^{\star}}^{H(3)}(\mu_0=m_t)$ as a piecewise function
using the expansions in the various limits. In Fig.~\ref{fig::c73l2hdm}(b) ,
the corresponding results for the coefficient $C_{7,A_d
A_u^{\star}}^{H(3)}(\mu_0=m_t)$ are plotted showing the same features,
however, with a smaller overlap of the $\mhplus\gg m_t$ and $\mhplus\approx
m_t$ curves. For the phenomenological analysis in the next Section, we define
\begin{eqnarray}
  C_{7,X}^{H(3)} = \left\{
    \begin{array}{ccc}
      C_{7,X}^{H(3)}(r\to\infty) 
      & \mbox{for} & \mhplus < 50~\mbox{GeV}\\
      C_{7,X}^{H(3)}(r\to1)                  
      &            & 50~\mbox{GeV} \le \mhplus < M_{7,X}\\
      C_{7,X}^{H(3)}(r\to0)                  
      &            & \mhplus \ge M_{7,X}
    \end{array}
  \right.
  \,,
  \label{eq::c7AuAuCpiecewise}
\end{eqnarray}
with $M_{7,A_uA_u^\star}=520\,$GeV and $M_{7,A_dA_u^\star}=400\,$GeV.

Analogous results for $C_8^{H(3)}(\mu_0=m_t)$ are shown in
Fig.~\ref{fig::c83l2hdm}. We observe the same pattern as for 
$C_7^{H(3)}(\mu_0=m_t)$, which leads us to define
\begin{eqnarray}
  C_{8,X}^{H(3)} = \left\{
    \begin{array}{ccc}
      C_{8,X}^{H(3)}(r\to\infty) 
      & \mbox{for} & \mhplus < 50~\mbox{GeV}\\
      C_{8,X}^{H(3)}(r\to1)                  
      &            & 50~\mbox{GeV} \le \mhplus < M_{8,X}\\
      C_{8,X}^{H(3)}(r\to0)                  
      &            & \mhplus \ge M_{8,X}
    \end{array}
    \right.
    \,,
   \label{eq::c8AuAuCpiecewise}
\end{eqnarray}
with $M_{8,A_uA_u^\star}=600\,$GeV and $M_{8,A_dA_u^\star}=520\,$GeV.

The {\tt Mathematica} file in Ref.~\cite{progdata} contains
the definitions~(\ref{eq::c7AuAuCpiecewise}) and~(\ref{eq::c8AuAuCpiecewise}),
which allows for convenient numerical evaluation of the 2HDM contributions to
$C_7$ and $C_8$. The updated SM results are included there, too.



\section{\label{sec::BR2hdm}${\cal B}(\bar{B}\to X_s \gamma)$ in the 2HDM to
  NNLO}

The framework for our numerical analysis is based on Ref.~\cite{Misiak:2006ab}
where explicit results for the effective-theory description of ${\cal
B}(\bar{B}\to X_s \gamma)$ have been provided up to the NNLO.  While
the Wilson coefficients are known in a complete manner at this order, non-BLM
NNLO corrections to the charm-quark-mass-dependent matrix elements (on-shell
amplitudes) have been evaluated only in the large $m_c$ limit and extrapolated
to the physical region.
\begin{figure}[t]
\begin{center}
  \begin{tabular}{cc}
    \includegraphics[width=0.48\textwidth]{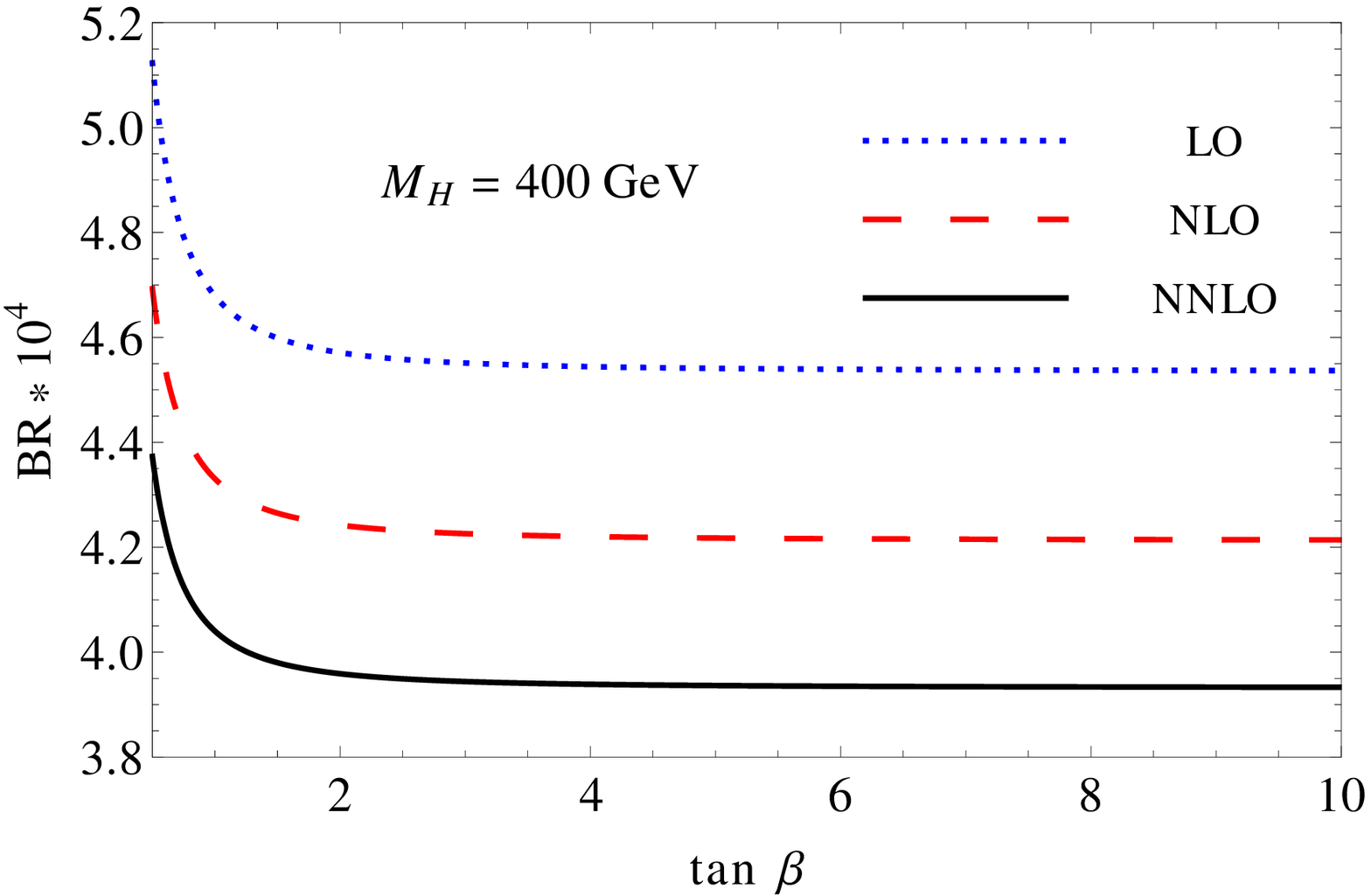} &
    \includegraphics[width=0.48\textwidth]{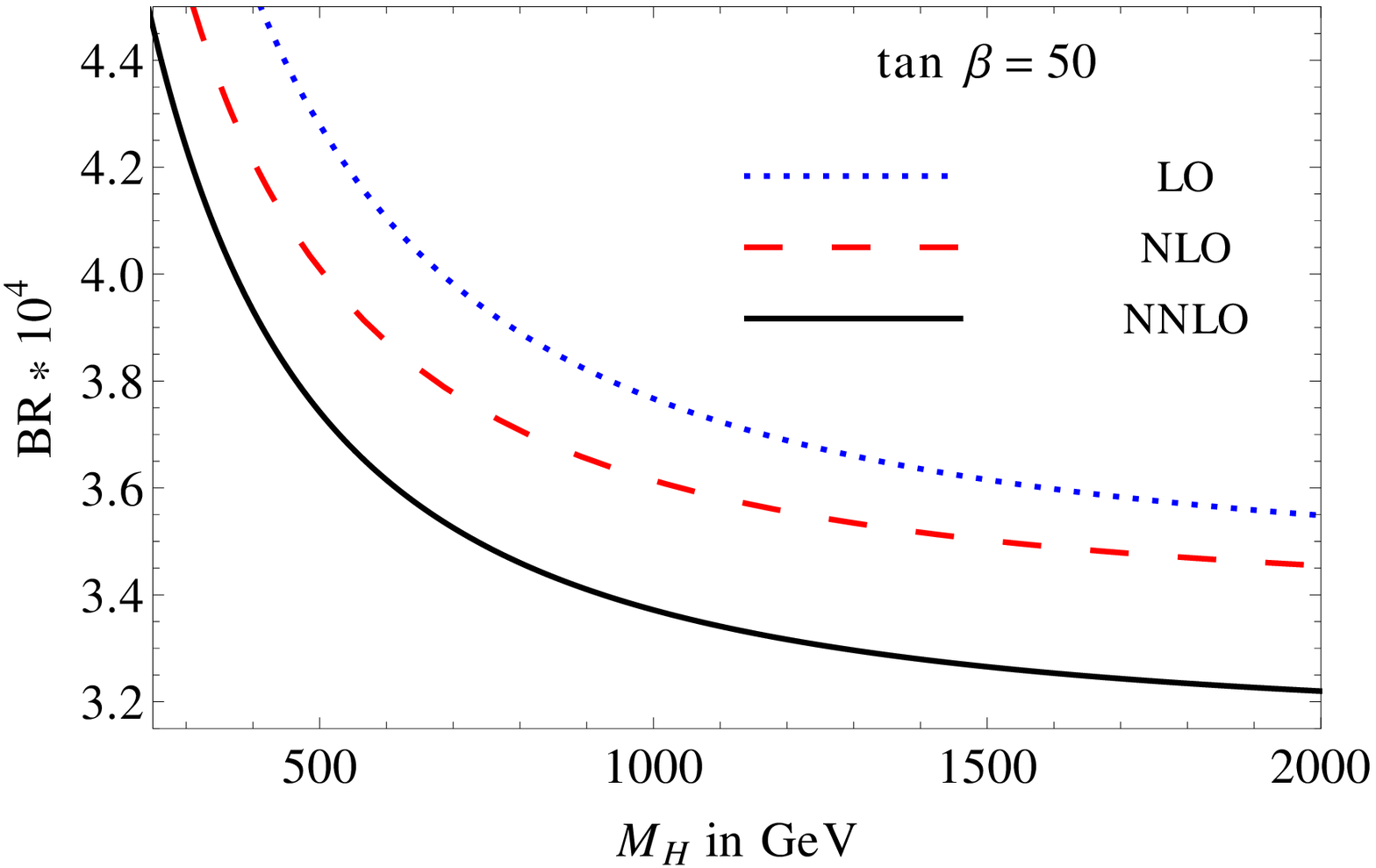}
    \\
    (a) & (b)
  \end{tabular}
    \caption{
      \label{fig::BRtanbeta}
      ${\cal B}(\bar B\to X_s\gamma)$ in the 2HDM type-II as a function of $\tan\beta$ 
        for $\mhplus = 400\,$GeV (left plot), and as a function of $\mhplus$ for $\tan\beta=50$ (right plot).
        Dotted, dashed and solid lines show central values (without uncertainties) 
        of the LO, NLO and NNLO results, respectively.}
  \end{center}
\end{figure}

Predictions within the 2HDM to be discussed below are obtained along the
same algorithm and using the complete NNLO matching conditions from the
previous Section. However, two-loop purely electroweak corrections to the
matching are included in the SM part only, as they remain unknown in the
2HDM. One should keep in mind that such electroweak corrections and
our new NNLO QCD matching ones may be of comparable size.

In the following, we shall discuss results for the 2HDMs of type-I and~II that
have been introduced in Eqs.~(\ref{eq::typeI}) and~(\ref{eq::typeII}). Most of
the input parameters are adopted from Ref.~\cite{Misiak:2006ab}, except for
the strong coupling constant and the top quark mass for which we use the most
up-to-date values that are given
by~\cite{Aaltonen:2012ra,Beringer:2012xxx,Bethke:2012zza}.\footnote{
Conservatively, we use ``0.0014'' as the uncertainty for $\alpha_s$ instead of
``0.0007''\cite{Beringer:2012xxx,Bethke:2012zza}.  This is motivated by the current
tension in several precision determinations of $\alpha_s$ (see discussion in
Ref.~\cite{Bethke:2011tr}).}
\begin{eqnarray}
  \alpha_s(M_Z) &=& 0.1184 \pm 0.0014\,,\nonumber\\
  M_t &=& (173.18 \pm 0.56_{\rm stat} \pm 0.75_{\rm syst})\,\mbox{GeV}\,.
\end{eqnarray}
The corresponding $\overline{\rm MS}$ top quark mass equals $m_t(m_t) =
163.5\,$GeV using three-loop accuracy in
QCD~\cite{Chetyrkin:1999qi,Melnikov:2000qh} and neglecting
electroweak effects. As in Ref.~\cite{Misiak:2006ab}, our default value of the
photon energy cut is $E_0 = 1.6\,$GeV.  Furthermore, if not stated otherwise,
we choose $\mu_0 = 160\,$GeV, $\mu_b = 2.5\,$GeV and $\mu_c = 1.5\,$GeV for the
renormalization scales, where $\mu_0$ is the matching scale, $\mu_b$ is the
scale at which on-shell matrix elements in the effective theory are evaluated,
and $\mu_c$ is the charm quark mass renormalization scale.

In a first step, let us discuss the branching ratio dependence on
$\tan\beta$. In Fig.~\ref{fig::BRtanbeta}(a), we choose 
$\mhplus=400\,$GeV and show ${\cal B}(\bar B\to X_s\gamma)$ in the 2HDM
type-II for $0.5\le\tan\beta\le10$. The solid curve describes the NNLO
result, while the dotted and dashed ones show the LO and NLO central values
for comparison. One observes strong dependence for $\tan\beta<2$ and a
nearly $\tan\beta$-independent result for $\tan\beta>2$. Actually, from
$\tan\beta=10$ to $\tan\beta=50$, the branching ratio changes only by
$0.03$\%.  In the following, $\tan\beta=50$ is going to be our
default value for the type-II model; choosing $\tan\beta<2$ would strengthen 
the lower limit on $\mhplus$.

In Fig.~\ref{fig::BRtanbeta}(b), we show ${\cal B}(\bar B\to
X_s\gamma)$ in the same model with $\tan\beta=50$ as a function of 
$\mhplus$. As expected, for large values of $\mhplus$, the 2HDM result
approaches the SM one that overlaps with the bottom frame of the plot in the
NNLO case. For $\mhplus=300\,$GeV the NNLO curve overshoots
the SM prediction by about 35\%, while the effect decreases to
around 2\% at $\mhplus=2$~TeV.
\begin{figure}[t]
\begin{center}
  \begin{tabular}{cc}
    \includegraphics[width=0.48\textwidth]{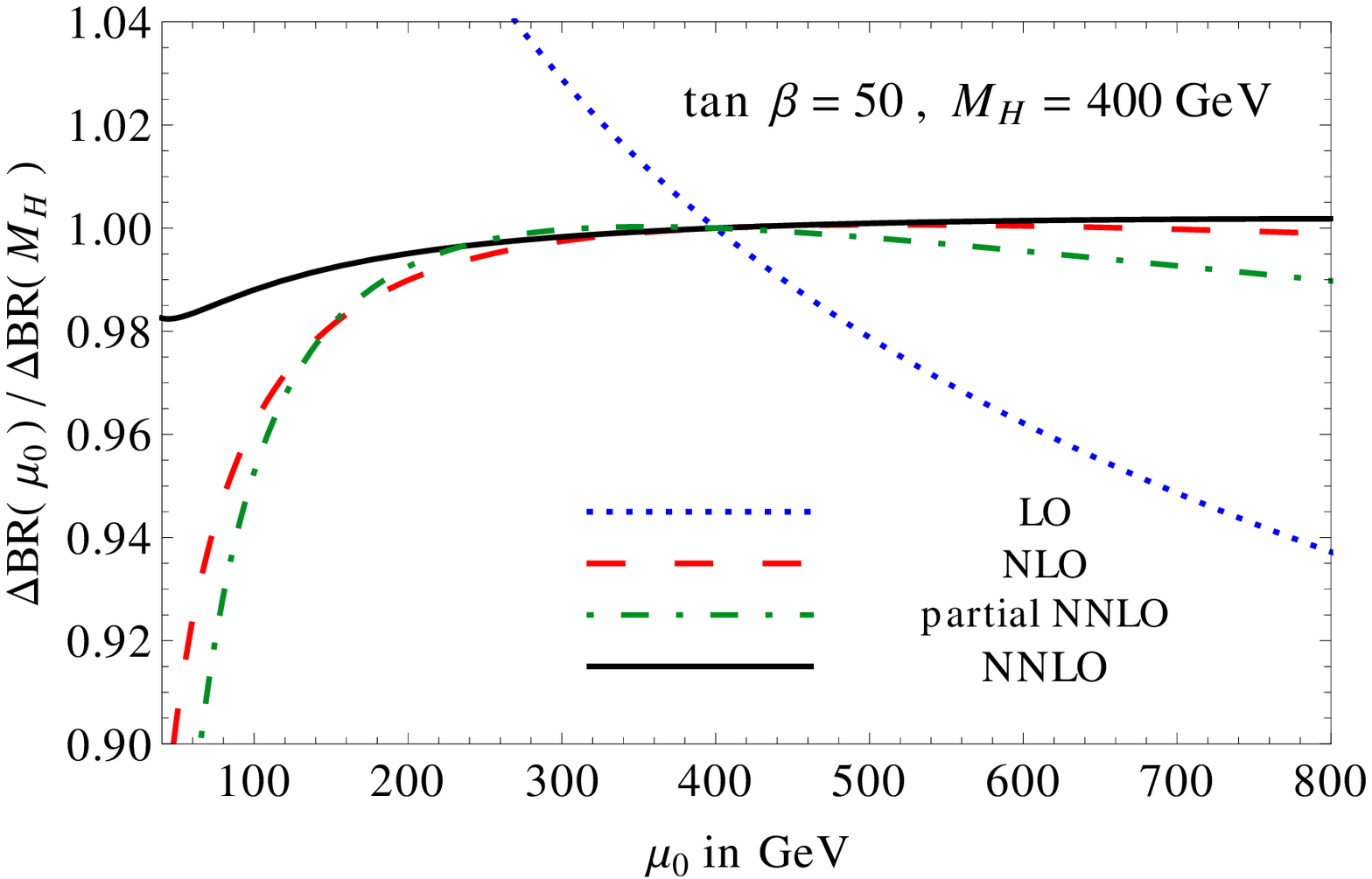} &
    \includegraphics[width=0.48\textwidth]{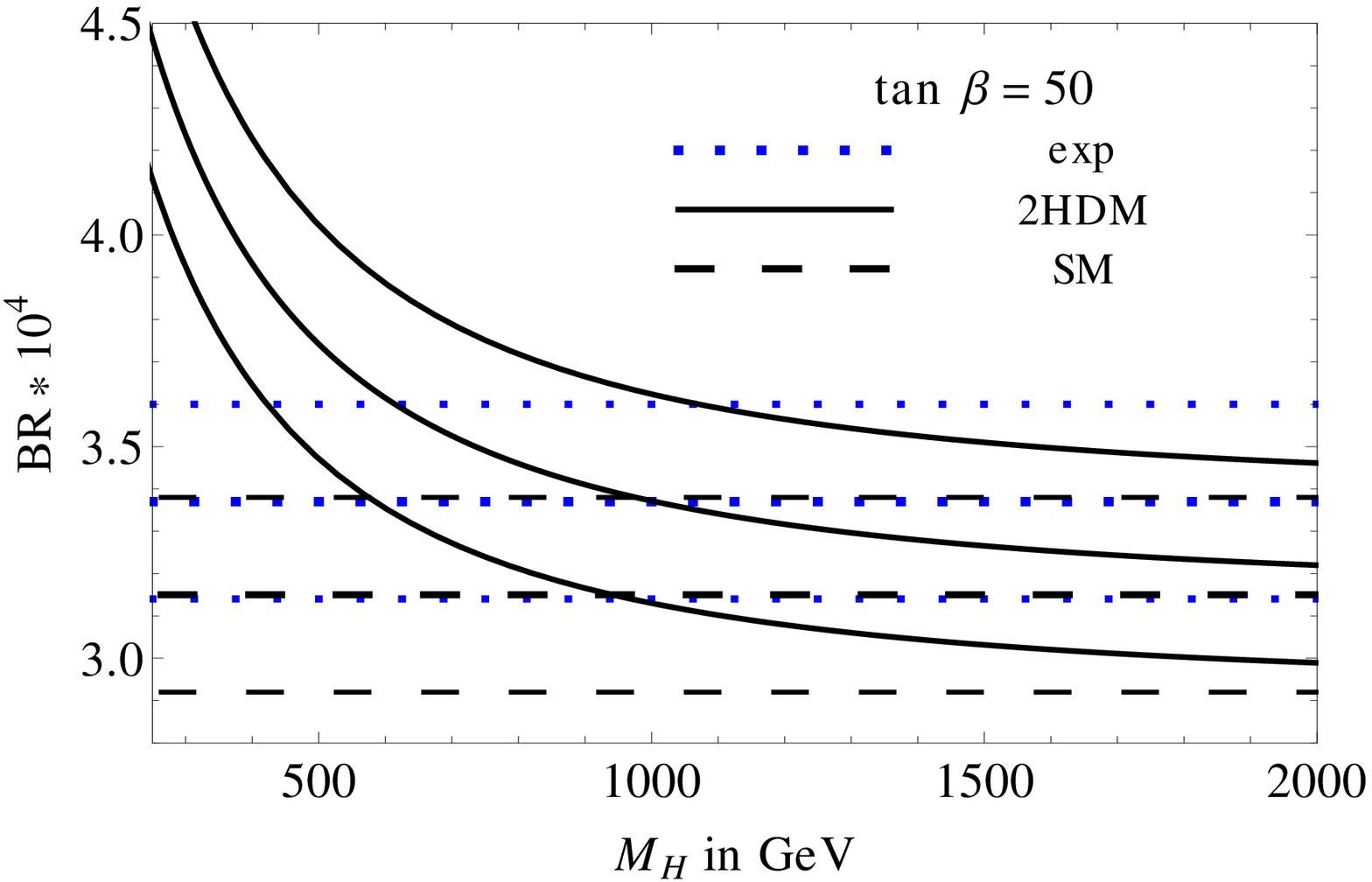}
    \\
    (a) & (b)
  \end{tabular}
    \caption{
      \label{fig::type2}
      A closer look at the 2HDM type-II results. 
      Left:  $\Delta{\cal B}(\mu_0)/\Delta{\cal B}(\mu_0=\mhplus)$ as a function of $\mu_0$ 
             for $\tan\beta=50$ and $\mhplus = 400\,$GeV at the 
             LO (dotted), NLO (dashed) and NNLO (solid). Dash-dotted lines
             correspond to the partial NNLO result which has been used in
             Ref.~\cite{Misiak:2006zs}. 
      Right: ${\cal B}(\bar B\to X_s\gamma)$ as a function of $\mhplus$ 
             for $\tan\beta=50$. Middle lines show the central values, while the upper and 
             lower ones are shifted by $\pm 1\sigma$. Solid and dashed lines correspond 
             to the NNLO 2HDM and SM predictions, respectively. Dotted curves 
             represent the experimental average in Eq.~(\ref{exp.aver}).}
  \end{center}
\end{figure}

The main effect of our new three-loop terms is in reducing
$\mu_0$-dependence of the decay rate and, in consequence, stabilizing the
lower bound on $\mhplus$. This is illustrated in Fig.~\ref{fig::type2}(a) 
where the charged Higgs contribution to the branching ratio
$\Delta{\cal B} \equiv {\cal B}_{\rm 2HDM} - {\cal B}_{\rm SM}$ 
is plotted as a function of $\mu_0$, while normalized to its own value at
$\mu_0 = \mhplus = 400\,$GeV. Apart from the LO (dotted), NLO (dashed),
NNLO (solid) curves, we also present the partial NNLO (dash-dotted)
line that corresponds to the approach of Ref.~\cite{Misiak:2006zs}.  Our
calculation differs from the latter one precisely by including the 2HDM
contributions to the NNLO matching. One observes a clear reduction of 
$\mu_0$-dependence when including higher order corrections. Whereas
the partial NNLO result for the considered ratio varies by 
more than $6.6\%$
when $\mu_0$ is varied in the $\left[ 80~\mbox{GeV},~ 2\mhplus\right]$ range, 
the corresponding variation of our present result with full NNLO matching 
remains below $1.6\%$. 
The overall size of $\Delta{\cal B}(\mu_0)$ amounts to
around $25\%$
of ${\cal B}(\bar B\to X_s\gamma)_{\rm SM}$ in the considered case.

In order to determine a lower bound on $\mhplus$ we follow the approach of
Ref.~\cite{Misiak:2006zs}, combining the experimental and theoretical
uncertainties in quadrature.  A one-sided 95\%~C.L. (99\%~C.L.) bound is
obtained for the lowest value of $\mhplus$ for which the difference between
experimental and theoretical central values is 1.645 (2.326) times larger than
the total uncertainty.

The theory uncertainty consists of four contributions that we take over
from Ref.~\cite{Misiak:2006ab}. Let us briefly comment on each of them:
\begin{itemize}
\item In Ref.~\cite{Misiak:2006ab}, the non-perturbative uncertainty has been
  estimated to $\pm 5\%$, which has been confirmed by the detailed
  investigation of Ref.~\cite{Benzke:2010js}.  We adopt this uncertainty for
  all the considered values of $\mhplus$.
\item The charm quark mass dependence of the operator matrix elements is only
  partly known. Thus, an interpolation between the large-$m_c$
  results~\cite{Misiak:2010sk} and reasonable assumptions for ${\cal B}(\bar
  B\to X_s\gamma)$ at $m_c=0$ has to be performed. This uncertainty has been
  estimated in~\cite{Misiak:2006ab} to $\pm 3\%$, which we again assume to be
  $\mhplus$-independent.

\item The total parametric error is obtained by combining all the partial ones
    in quadrature.\footnote{A correlation between the phase space factor $C$
    and $m_c$ is taken into account as described in Appendix~A of
    Ref.~\cite{Misiak:2006ab}.}  It amounts to around $2\div 3\%$, however,
    computed from scratch for each value of $\mhplus$.
\item An estimate of higher order corrections is obtained by varying the
  renormalization scales in the ranges $80~\mbox{GeV} \le \mu_0 \le
  \mbox{max}\{2\mhplus,320~\mbox{GeV}\}$, $1.25~\mbox{GeV} \le \mu_b \le
  5~\mbox{GeV}$ and $1.224~\mbox{GeV} \le \mu_c \le M_b^{1S}=4.68~\mbox{GeV}$.
  This uncertainty is about  $3 \div4\%$ but again we compute it for each
  value of $\mhplus$.
\end{itemize}
Contrary to Refs.~\cite{Misiak:2006ab,Misiak:2006zs}, we work with
asymmetric uncertainties resulting from the parameter and renormalization
scale variation.
\begin{figure}[t]
  \begin{center}
    \begin{tabular}{cc}
      \includegraphics[width=0.48\textwidth]{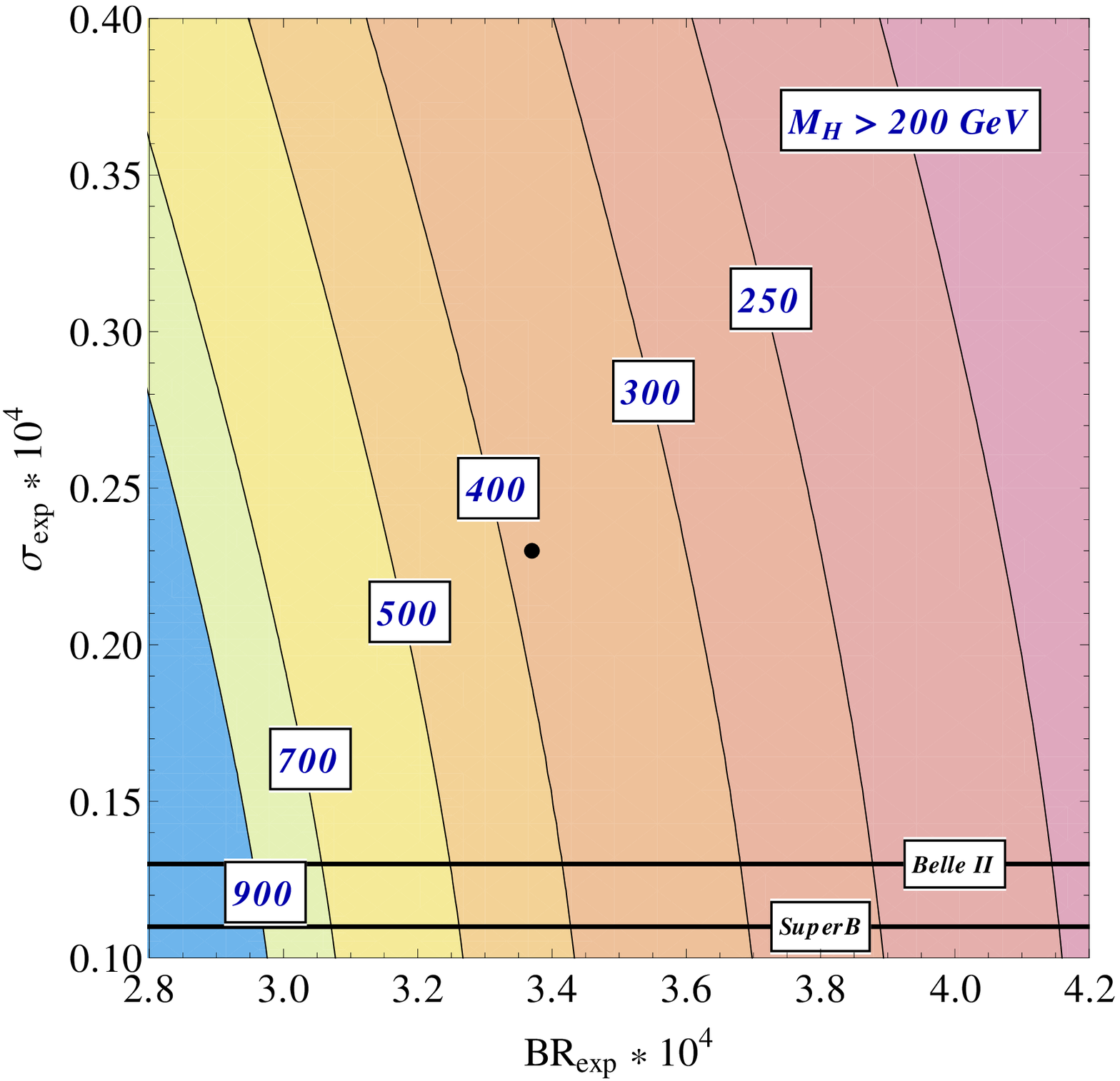}
      &
      \includegraphics[width=0.48\textwidth]{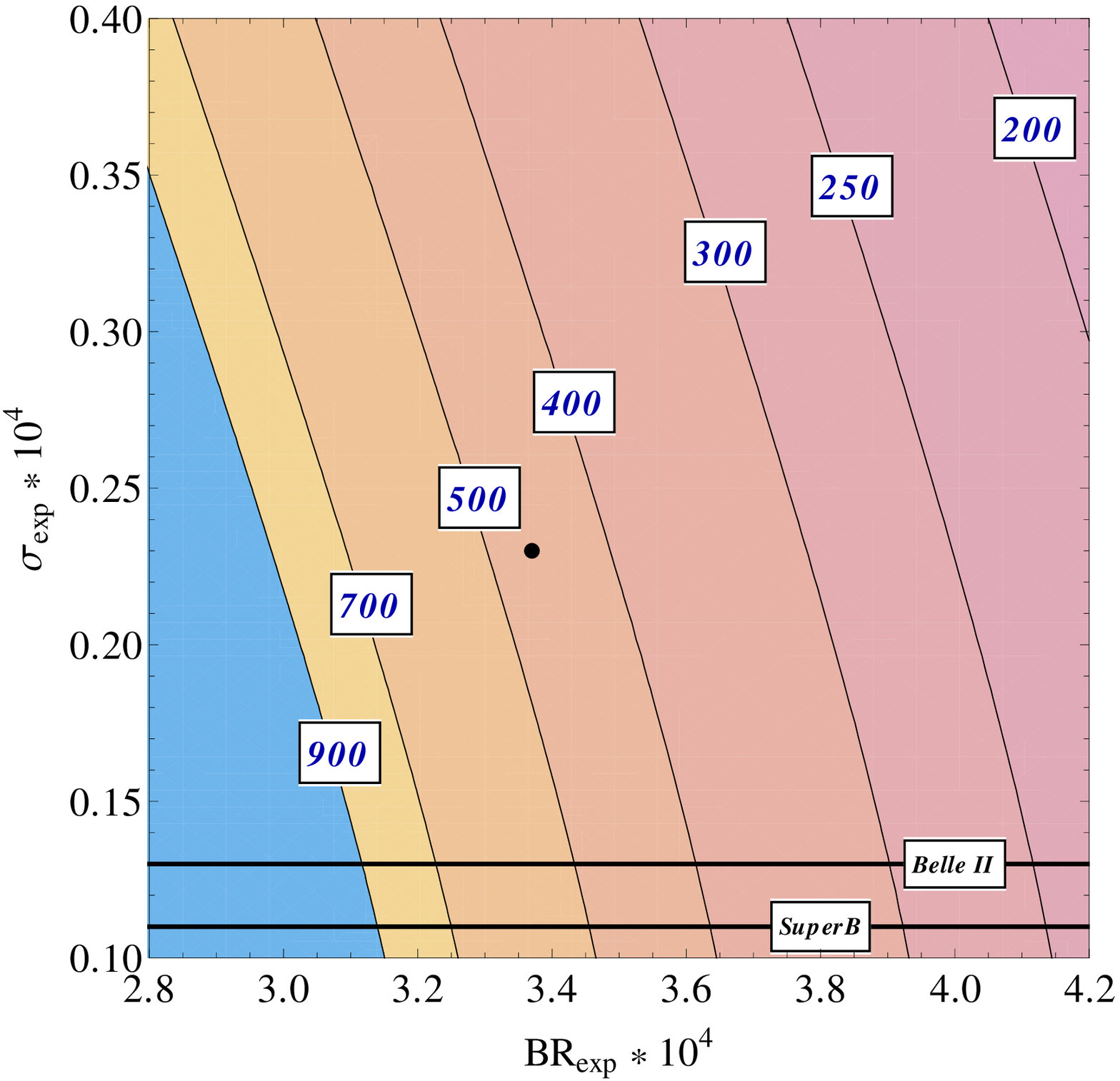}
      \\
      (a) & (b)
    \end{tabular}
    \caption{
      \label{fig::BRexpdeltaBRexp}
      Lower bounds on $\mhplus$ at the 95\% C.L. as a function of the
      experimentally determined branching ratio (abscissa) and the corresponding
      uncertainty (ordinate). The current theory uncertainty has been used
      in panel (a), while panel (b) presents a future projection with assumed
      reduction of the theory uncertainty by a factor of two.}
  \end{center}
\end{figure}

In Fig.~\ref{fig::type2}(b), the NNLO result for ${\cal B}(\bar B\to X_s\gamma)$
is shown as a function of $\mhplus$, together with an uncertainty band that
is obtained by adding all the errors in quadrature. The dotted 
($\mhplus$-independent) curves in Fig.~\ref{fig::type2}(b) correspond to the experimental
result in Eq.~(\ref{exp.aver}). For comparison, we also show the SM
prediction with the corresponding uncertainty as dashed lines.

From Fig.~\ref{fig::type2}(b) one can extract (using the procedure described
above) the following limits on $\mhplus$
in the 2HDM type-II:
\begin{eqnarray}
\mhplus &\ge& 380~\mbox{GeV} \quad \mbox{at } 95\%~\mbox{C.L.}\,,\nonumber\\
\mhplus &\ge& 289~\mbox{GeV} \quad \mbox{at } 99\%~\mbox{C.L.}\,.
\label{eq::mhplus_limit}
\end{eqnarray}
The above bounds replace the ones of Ref.~\cite{Misiak:2006zs} ($295$ and
$230\,$GeV, respectively). 
A considerable improvement of the bounds arises mostly due to the
shift in the experimental value of ${\cal B}(\bar{B}\to X_s \gamma)$
that is now smaller than the one used in Ref.~\cite{Misiak:2006zs}.
Our 95\%~C.L. limit is very close to the one
presented in Ref.~\cite{Stone:ICHEP2012} ($385\,$GeV) together with the new
experimental average (Eq.~(\ref{exp.aver})). On the other hand, it is
significantly stronger than the one in Ref.~\cite{Lees:2012ym} ($327\,$GeV)
that is based on the BABAR data alone. It is interesting to mention that
when the matching scale $\mu_0$ is varied between $80$ and $400$~GeV, the
lower limits vary by around $25$~GeV when our new three-loop 2HDM matching
contributions are {\em not} included. This gets reduced to around $7$~GeV only
after including the three-loop corrections, which demonstrates the stabilizing
effect of the full NNLO matching. On the other hand, for $\mu_0=160$~GeV
fixed, the correction strengthens the limit only slightly, by $5\div 6$~GeV.

The contour plots in Fig.~\ref{fig::BRexpdeltaBRexp} show the
95\% C.L. lower bounds on $\mhplus$ as functions of the
experimentally determined branching ratio and the corresponding uncertainty.
Black dots correspond to the result in
Eq.~(\ref{exp.aver}), while the two black lines at the bottom indicate the
projected uncertainty to be reached by Belle~II and
SuperB~\cite{Meadows:2011bk}.  The current theory uncertainty has been used in
Fig.~\ref{fig::BRexpdeltaBRexp}(a), while Fig.~\ref{fig::BRexpdeltaBRexp}(b)
presents a future projection with assumed reduction of the theory uncertainty
by a factor of two.
\begin{figure}[t]
  \begin{center}
    \begin{tabular}{cc}
      \includegraphics[width=0.48\textwidth]{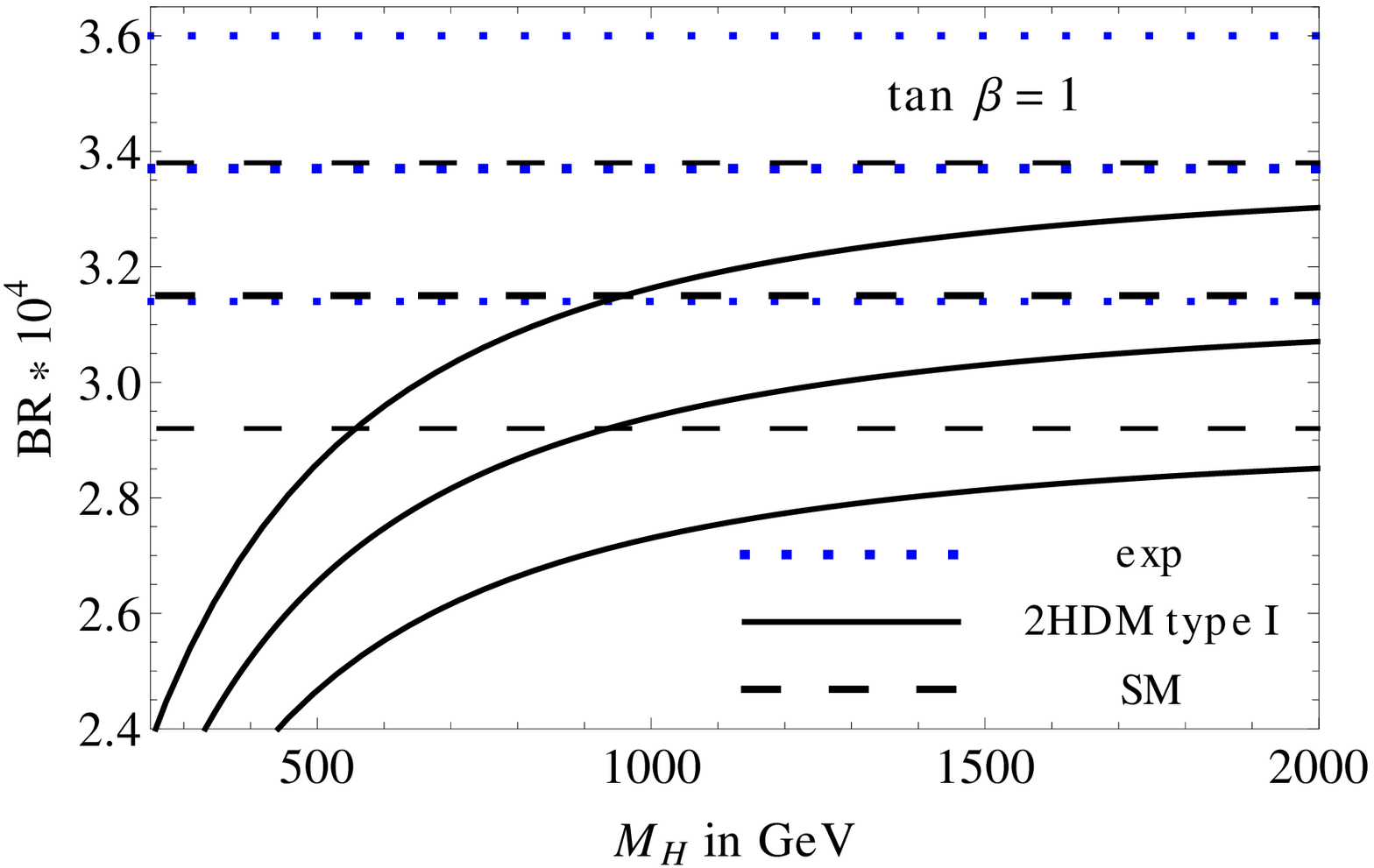}
      &
      \includegraphics[width=0.48\textwidth]{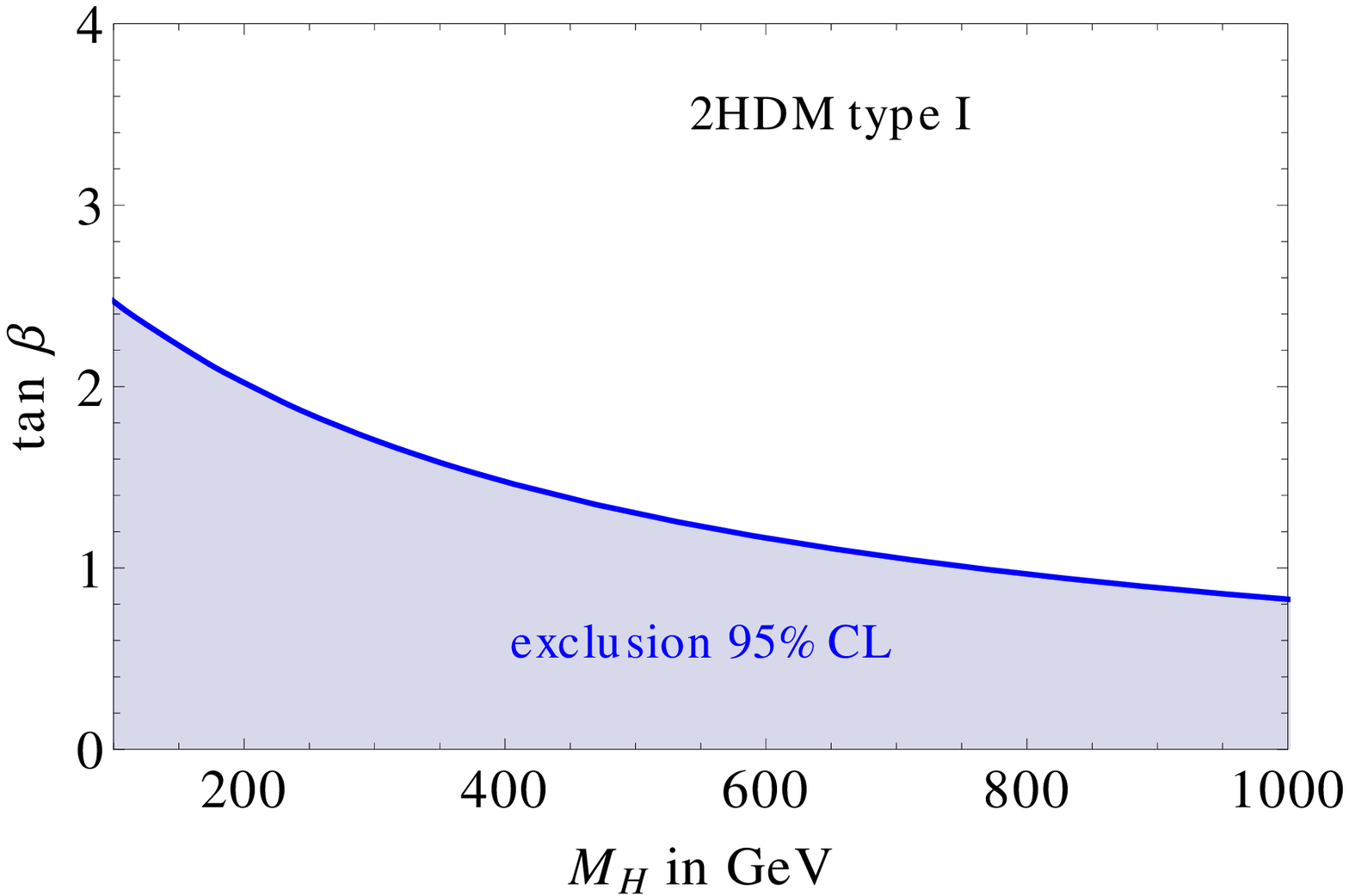}
      \\
      (a) & (b)
    \end{tabular}
    \caption{
      \label{fig::type1}
      Left: ${\cal B}(\bar B\to X_s\gamma)$ for 2HDM type-I with $\tan\beta=1$.
      Identification of the lines is the same as in Fig.~\ref{fig::type2}(b).
      Right: 95\% C.L. exclusion (shaded area) for 2HDM type-I in the plane
      of $\tan\beta$ and $\mhplus$.}
  \end{center}
\end{figure}

Let us now shortly discuss the 2HDM of type-I. In Fig.~\ref{fig::type1}(a), we
show the branching ratio for $\tan\beta=1$ as a function
of $\mhplus$, including the uncertainties. The meaning of the curves is the
same as in Fig.~\ref{fig::type2}(b). Contrary to the type-II model, the
branching ratio gets suppressed with respect to the SM, and the effect becomes
larger for lower values of $\mhplus$, which implies increased discrepancy with
respect to the experimental results.  Thus, it is possible to set a lower
bound on $\mhplus$. It is shown in Fig.~\ref{fig::type1}(b) where the shaded
area represents the part of the $\tan\beta$-$\mhplus$ plane that is excluded
at 95\% C.L.  Similar results based on the partial NNLO predictions and
the previous experimental average have been presented in
Ref.~\cite{Mahmoudi:2009zx}.


\section{Conclusions}\label{sec::conclusions}

Applying the method of expansion in mass ratios, we have evaluated
three-loop matching conditions for the dipole operators $P_7$ and $P_8$ in the
2HDM. In effect, Wilson coefficients of all the operators that matter for
$\bar{B}\to X_s \gamma$ in this model at the leading order in electroweak
interactions are now known to the NNLO accuracy in QCD.  This is true not only
at the matching scale $\mu_0$ but also at the low-energy scale $\mu_b$ because
the renormalization group evolution is the same as in the SM~\cite{Czakon:2006ss}.

The main effect of including the NNLO matching is a significant reduction of
$\mu_0$-dependence of the charged Higgs contribution to the $\bar{B}\to X_s
\gamma$ branching ratio. It is particularly transparent when considering a
lower bound on $\mhplus$ in the type-II model. Before taking our correction
into account, the bound varies by around $25$~GeV when $\mu_0$ is
varied in a reasonable range $[80,400]\,{\rm GeV} \sim [\frac12
m_t,\mhplus]$. Including the correction reduces the variation by more than a
factor of 3.

With the updated experimental average, we find that the 95\%~C.L. (99\%~C.L.)
lower limit on $\mhplus$ amounts to 380~(289)~GeV in the 2HDM
type-II. This is a universal ($\tan\beta$-independent) bound that can only get
stronger when $\tan\beta$-dependence is taken into account. In practice,
noticeable modifications occur for $\tan\beta$ smaller than around 2.

In the 2HDM type-I, a 95\%~C.L. lower limit on $\mhplus$ from ${\cal
B}(\bar{B}\to X_s \gamma)$ can be derived for low $\tan\beta$ only, currently
for $\tan\beta < 2.5$ when $\mhplus$ is above the LEP
bound of around $80\,$GeV~\cite{Beringer:2012xxx}. In this case,
considerable reduction of $\mu_0$-dependence is observed, too.

With the semi-analytical results presented in this paper, constraints on the
2HDM at the NNLO level can easily be updated in the future, along with 
developments in the measurements and in calculations of the low-energy
matrix elements (that are identical in the SM and in the 2HDM). More precise
measurements are expected in a few years from Belle-II~\cite{Abe:2010sj} and
Super-B~\cite{O'Leary:2010af}.  On the theoretical side, the main challenges
are improvements in analyses of non-perturbative effects~\cite{Benzke:2010js}
together with perturbative calculations of the NNLO on-shell amplitudes beyond
the large-$m_c$ limit.  In the latter case, new results should become
available soon~\cite{Misiak:2010dz}.

 
\section*{Acknowledgements}

This work has been supported by the DFG through the SFB/TR~9 ``Computational
Particle Physics'' and the Graduiertenkolleg ``Elementarteilchenphysik bei
h\"ochster Energie und h\"ochster Pr\"azision''. M.M. acknowledges partial
support from the National Science Centre (Poland) research project, decision no
DEC-2011/01/B/ST2/00438, as well as from the DFG through the ``Mercator'' guest
professorship programme.


\section*{Note added} 
Shortly after our paper had been submitted, a new experimental value for the branching ratio
${\cal B}(\bar{B}\to X_s \gamma)|_{E_\gamma>1.6~\mbox{\tiny GeV}} =(3.43 \pm 0.22) \times 10^{-4}$
appeared at the HFAG web page~\cite{hfag}. If this result was used instead of
Eq.~(\ref{exp.aver}), our bounds in Eq.~(\ref{eq::mhplus_limit}) would read
\begin{eqnarray*}
\mhplus &\ge& 360~\mbox{GeV} \quad \mbox{at } 95\%~\mbox{C.L.}\,,
\\
\mhplus &\ge& 277~\mbox{GeV} \quad \mbox{at } 99\%~\mbox{C.L.}\,.
\end{eqnarray*}




\end{document}